\documentclass[longauth]{aa} 

\usepackage{graphicx}
\usepackage{txfonts}
\usepackage{soul}

\begin{document} 

   \title{CEERS: Forging the First Dust Grains in the Universe?}

   \subtitle{A Population of Galaxies with spectroscopically-derived Extremely Low Dust Attenuation (GELDA) at 4.0 $<$ z $\lesssim$ 11.4}

   \author{Denis Burgarella
          \inst{1}
          \and
          V\'eronique Buat\inst{1}
          \and
          Patrice Theul\'e\inst{1}
          \and
          Jorge Zavala\inst{2, 35}
        \and
          Mark Dickinson\inst{3}
          \and
          Pablo Arrabal Haro\inst{3}
          \and
          Micaela B. Bagley\inst{4}
          \and
          M\'ed\'eric Boquien\inst{5}
          \and
          Nikko Cleri\inst{6,7,8}
         \and
          Tim Dewachter\inst{1}
          \and
          Henry C. Ferguson\inst{9}
         \and
          Vital Fern\`andez\inst{10}
         \and
          Steven L. Finkelstein\inst{11}
         \and
          Eric Gawiser\inst{12}
         \and
          Andrea Grazian\inst{13}
         \and
          Norman Grogin\inst{9}
         \and
          Benne W. Holwerda\inst{14}
         \and
          Jeyhan S. Kartaltepe\inst{15}
         \and
          Lisa Kewley\inst{16}
         \and
          Allison Kirkpatrick\inst{17}
         \and
          Dale Kocevski\inst{18}
         \and
          Anton M. Koekemoer\inst{9}
         \and
          Arianna Long\inst{4}
         \and
          Jennifer Lotz\inst{9}
         \and
          Ray A. Lucas\inst{9}
         \and
          Bahram Mobasher\inst{19}
         \and
          Casey Papovich\inst{20,21}\
         \and
          Pablo G. P\'erez-Gonz\`alez\inst{22}
         \and
          Nor Pirzkal\inst{23}
         \and
          Swara Ravindranath\inst{24,25}
         \and
          Giulia Rodighiero\inst{26,27}
         \and
          Yannick Roehlly\inst{1}
         \and
          Caitlin Rose\inst{15}
         \and
          Lise-Marie Seill\'e\inst{1}
         \and
          Rachel Somerville\inst{28}
         \and
          Steve Wilkins\inst{29,30}
         \and
          Guang Yang\inst{31}
         \and
          L. Y. Aaron Yung\inst{9}
          }

   \institute{Aix Marseille Université, CNRS, CNES, LAM, Marseille, France\
              \email{denis.burgarella@lam.fr}
     \and
   National Astronomical Observatory of Japan, 2-21-1 Osawa, Mitaka, Tokyo 181-8588, Japan
     \and
   NSF's National Optical-Infrared Astronomy Research Laboratory, 950 N. Cherry Ave., Tucson, AZ 85719, USA
     \and
   Department of Astronomy, The University of Texas at Austin, Austin, TX, USA
     \and
   Université Côte d’Azur, Observatoire de la Côte d’Azur, CNRS, Laboratoire Lagrange, 06000, Nice, France
     \and
   Department of Astronomy and Astrophysics, The Pennsylvania State University, University Park, PA 16802, USA
     \and
   Institute for Computational and Data Sciences, The Pennsylvania State University, University Park, PA 16802, USA
     \and
   Institute for Gravitation and the Cosmos, The Pennsylvania State University, University Park, PA 16802, USA
     \and
   Space Telescope Science Institute, Baltimore, MD, USA
     \and
   nstituto de Investigación Multidisciplinar en Ciencia y Tecnología, Universidad de La Serena, Raul Bitràn 1305, La Serena 2204000, Chile
     \and
   epartment of Astronomy, The University of Texas at Austin, Austin, TX, USA
     \and
   INAF - Osservatorio Astronomico di Roma, via di Frascati 33, 00078 Monte Porzio Catone, Italy\
     \and
   Department of Physics and Astronomy, Rutgers, the State University of New Jersey, Piscataway, NJ 08854, USA
     \and
   INAF--Osservatorio Astronomico di Padova, Vicolo dell'Osservatorio 5, I-35122, Padova, Italy
     \and
   Physics \& Astronomy Department, University of Louisville, 40292 KY, Louisville, USA
     \and
   Laboratory for Multiwavelength Astrophysics, School of Physics and Astronomy, Rochester Institute of Technology, 84 Lomb Memorial Drive, Rochester, NY 14623, USA
     \and
   Center for Astrophysics, Harvard \& Smithsonian, 60 Garden Street, Cambridge, MA 02138, USA
     \and
   Department of Physics and Astronomy, University of Kansas, Lawrence, KS 66045, USA
     \and
   Department of Physics and Astronomy, Colby College, Waterville, ME 04901, USA
     \and
   Department of Physics and Astronomy, University of California, 900 University Ave, Riverside, CA 92521, USA
     \and
   Department of Physics and Astronomy, Texas A\&M University, College Station, TX, 77843-4242 USA
     \and
   George P. and Cynthia Woods Mitchell Institute for Fundamental Physics and Astronomy, Texas A\&M University, College Station, TX, 77843-4242 USA
     \and
   Centro de Astrobiologia (CAB), CSIC-INTA, Ctra. de Ajalvir km 4, Torrejon de Ardoz, E-28850, Madrid, Spain
     \and
   ESA/AURA Space Telescope Science Institute
     \and
   Astrophysics Science Division, NASA Goddard Space Flight Center, 8800 Greenbelt Road, Greenbelt, MD 20771, USA
     \and
   enter for Research and Exploration in Space Science and Technology II, Department of Physics, Catholic University of America, 620 Michigan Ave N.E., Washington DC 20064, USA
     \and
   Department of Physics and Astronomy, Università degli Studi di Padova, Vicolo dell’Osservatorio 3, I-35122, Padova, Italy
     \and
   INAF - Osservatorio Astronomico di Padova, Vicolo dell’Osservatorio 5, I-35122, Padova
     \and
   Center for Computational Astrophysics, Flatiron Institute, 162 5th Avenue, New York, NY, 10010, USA
     \and
   Astronomy Centre, University of Sussex, Falmer, Brighton BN1 9QH, UK
     \and
   Institute of Space Sciences and Astronomy, University of Malta, Msida MSD 2080, Malta
     \and
   Nanjing Institute of Astronomical Optics and Technology, Nanjing 210042, China
     \and
   University of Massachusetts Amherst, 710 North Pleasant Street, Amherst, MA 01003-9305, USA.\\
             }

   \date{Accepted May 2025}

  \abstract
     {}
  {This paper investigates the coevolution of metals and dust for 173 galaxies at 4.0$<$z$\leq$11.4 spectroscopically observed by JWST/NIRSpec in the CEERS project. More specifically, we want to study and analyze the properties of a sample of galaxies that show an extremely low dust attenuation and try to understand the possible physical processes at play in these galaxies.}
   {We develop a new version of the CIGALE code that accepts spectroscopic and photometric data. From a statistical comparison of the observations with the modeled spectra, we derive a set of physical parameters that allow us to constrain the above physical processes.}
   {Our analysis reveals a population of 49 extremely low dust attenuation galaxies (GELDAs) consistent with A$_{FUV}$ = 0.0 within 2$\sigma_{A\_FUV}$ and M$_{star}$ $<$ 10$^9$M$_\odot$. After stacking the spectra of the 49 GELDAs to increase the signal-to-noise ratio, we measure a very blue UV slope $\beta_{FUV}$ = -2.451 $\pm$ 0.066, and a Balmer decrement H$\alpha$/H$\beta$=2.932$\pm$0.660 without underlying absorption and consistent with no dust attenuation and Case B assuming an underlying absorption of 2.5 \%. Furthermore, the proportion of GELDA is much higher at z $>$ 8.8 (83.3 \% of the total sample) than at z $<$ 8.8 (26.3 \% of the total sample). This suggests that GELDAs become dominant in the early Universe.
   Assuming a prior far-infrared dust spectrum from the ALPINE sample, we perform an analysis of the properties of this galaxy population. The trends observed in the M$_{dust}$ vs. M$_{star}$ diagram feature an upper and a lower sequence linked by objects that can be transitional. 
   A comparison with models suggests that we might observe a critical transition at M$_{star}$ $\approx$ 10$^{8.5}$ M$\odot$, corresponding to a critical metallicity of Z$_{crit}$ = 12 + $\log_{10}$(O/H) $\approx$ 7.60 (i.e., Z/Z$\odot$ $\approx$ 0.1). At this point, galaxies transition from being dominated by stellar dust production (mainly from supernovae) to grain growth through gas–dust accretion in the ISM. The observational critical metallicity Z$_{crit}$ derived in this paper is in good agreement with predictions from theoretical models for the onset of efficient grain growth.
   Furthermore, the mean gas mass fraction of our entire sample at 4.0$<$z$<$11.4 is very high: f$_{gas}\gtrsim0.9$. All of our galaxies, including GELDAs at all redshifts, contain a large amount of gas that was not expelled from the galaxies. Finally, the small size of the galaxies combined with the mass of gas lead to very high surface gas densities, which put our sample below high-redshift sub-millimeter galaxies, at relatively low star formation efficiency. The population of high-redshift GELDAs would provide us with a natural and inherent explanation for the origin of the apparent tension between observations and theoretical models in the number density of bright galaxies at z $\gtrsim$ 9.}
   {}


   \maketitle

\section{Introduction}

After the big bang, nucleosynthesis started in the first stellar population formed in the Universe: population III stars (pop. III), and shortly after pop. II stars. Type II supernovae (SNae II) expelled metals much earlier than asymptotic giant branch (AGB) stars in the early Universe (\citealt{Valiante2009,Hirashita2014,DellAgli2019,Walter2020}), which then led to the formation of the first dust grains by coalescence of these metals (e.g. Schneider+24).
Dust affects ultraviolet (UV) and optical emissions through dust attenuation and reddening.  Dust IR emission is a major cooling agent. This energy warms dust and is re-emitted at infrared (IR) and sub-millimeter (sub-mm) wavelengths (Burgarella2005, Malek2018, Pozzi2021). 
Dust plays a critical role in the formation of low-mass stars by facilitating several key processes in the interstellar medium (ISM). Molecular hydrogen (H$_2$), which is essential for cloud collapse and subsequent star formation, does not form efficiently in the gas phase under typical ISM conditions. Instead, H$_2$ formation is catalyzed on the surfaces of dust grains, making dust indispensable to initiating star formation even at warm temperatures (\citealt{Grieco2023}). Once formed, these grains can act as seeds for further grain growth in the ISM, enhancing the dust mass available (\citealt{Zhukovska2018, Asano2013}). Furthermore, collisions between gas and dust grains enable efficient gas cooling, particularly at high densities (n$_H$$\sim$10$^{12}$ cm$^{-2}$), which promotes fragmentation of the cloud and the formation of low-mass stars. These low-mass stars may represent a transition population between Population III and Population II stars.

One of the main results from JWST's first years of observation is an unpredicted excess of UV-luminous galaxies at z$>$10 compared to HST-calibrated models (\citealt{Finkelstein2023, Finkelstein2024, Naidu2022, Casey2024}. The galaxies present far-UV (FUV) absolute magnitudes, -21 $\lesssim$ M$_{UV}$ $\lesssim$ -19, very blue FUV spectral slopes, $\beta_{FUV}$ $\lesssim$ -2.2, small effective radii, r$_e$ $\sim$ 200 - 400 pc) and stellar masses,  M$_{star}$ $\sim$ 10$^9$ M$_\odot$ (\citealt{Atek2023, Ferrara2025}) that are similar to ours, especially the highest redshift galaxies.

Several potential causes have been proposed. We cannot present an exhaustive inspection in this paper, and we suggest following some of the tracks opened in the next paragraphs for a full and detailed review. 

In brief, a first one (\citealt{Feldmann2025}) is proposed that relatively high star formation efficiency (SFE) in the early Universe is a natural outcome of the baryonic processes encoded in the FIRE-2 model (\citealt{Hopkins2018}) because the shallower slope of their SFE-M$_{halo}$ relation at $9 < \log_{10}(M_{halo}/M_\odot) < 11$ leads to an increase of the contribution from the more numerous lower mass haloes, and thus an increase of the observed abundances of bright galaxies at z $>$ 10.

Another proposed hypothesis is that this higher SFE could be due to feedback-free starbursts (FFBs) where the SFE could reach a maximum SFE = 0.2 - 1.0 in the FFB regime (\citealt{Li2024}), to the formation of Pop. III stars (e.g. \citealt{Yung2024} or dark stars with masses $\gtrsim 10^3$ M$_\odot$, which could be fueled by heating from dark matter in the first dark matter halos or minihalos (\citealt{Ilie2023, Lei2025}). The effect could be due to a top-heavy initial mass function (IMF): an increase in the characteristic stellar mass of a top-heavy IMF would add up massive stars, which in turn would produce more UV light (e.g. \citealt{Zackrisson2011, Harikane2023, Hutter2025, Jeong2025}). 

It could also be due to an increased stochasticity exposed through dispersion in the relation between galaxy UV magnitude (M$_{UV}$) and halo mass is also envisaged to explain the UV-bright overabundance: bursty star formation histories (SFHs) have been measured (e.g., \citealt{Cole2025}) via the scatter in the star-forming main sequence (MS). However, if such stochasticity in star formation can certainly help, stochasticity alone might not be enough (\citealt{Yung2024, Finkelstein2024}).

Finally, an origin related to dust, either as the only process or combined with another is certainly quite compelling: either a low dust attenuation and/or a specific dust-star geometries directly leading to unobscured young stars.

The origin of this unexpected excess has not yet been fully understood. In this paper, we advocate another possible dust-related origin that is supported by observations.

In the last decade, we have detected dusty galaxies at z$>$4 (\citealt{Vieira2013, Hezaveh2015, Watson2015, Laporte2017, Fudamoto2017, Strandet2017, Tamura2019, Sommovigo2022a, Algera2024a, Algera2024b, Witstok2023, Zavala2023, Valentino2024}) with large dust masses (\citealt{Pozzi2021, Akins2023}) that cannot be explained by models. SNae and AGB stars could scarcely be at the origin of such a large dust mass, especially if we account for the reverse shock resulting from the expanding SN blast wave in the ISM (\citealt{Lesniewska2019}). To explain such a large dust mass, we could assume an important role of dust mass growth in the ISM. 

Models of dust formation, both semi-analytical \citep[e.g.,][]{Popping2017, Vijayan2019, Triani2020, Dayal2022, Mauerhofer2023} and cosmological \citep[e.g.,][]{Graziani2020, Lewis2023, DiCesare2023, Choban2024} predict a critical metallicity in the ISM above which grain growth via gas–dust accretion becomes efficient, marking the transition from dust production dominated by stellar sources (supernovae and AGB stars) to accretion-driven dust growth in the ISM. While this transition is supported by local-Universe observations \citep{RemyRuyer2014, DeVis2019}, it remains undetected at high redshift.

\cite{Ferrara2022} suggest that dust could have been efficiently ejected during the very first phases of galaxy build-up. This hypothesis is supported by theoretical predictions that high-redshift galaxies are extremely bursty and very small (e.g. \citealt{Sun2023, Choban2024}). In this case, we would only detect the almost dust-free objects that remain. However, both star/ISM dust geometry and galactic dust ejections would produce the same low dust attenuations and thus bright UV luminosities. Other discriminant observables such as the gas mass fraction and the metallicity should be explored to study this possible degeneracy.

This paper derives new constraints on the ISM at high redshift from the JWST/CEERS  project\footnote{The Cosmic Evolution Early Release Science Survey, https://ceers.github.io/}, featuring NIRCAM, NIRSpec, plus ancillary data from SCUBA-2 (\citealt{Zavala2018}),  and NOEMA (\citealt{Fudamoto2017}) with the help of a new version of the CIGALE code that accepts spectrophotometric data. More technical details on the new version of CIGALE are provided in the appendix~\ref{appendix:CIGALE}.

\section{Observations}
\label{sect.observations}
\subsection{The NIRSpec prism spectroscopic sample}
\label{sect.NIRSpec.sample}
\subsubsection{The origin of the sample}
\label{sect.origin.sample}

\begin{figure}
    \begin{centering}
	\includegraphics[width=0.8\columnwidth]{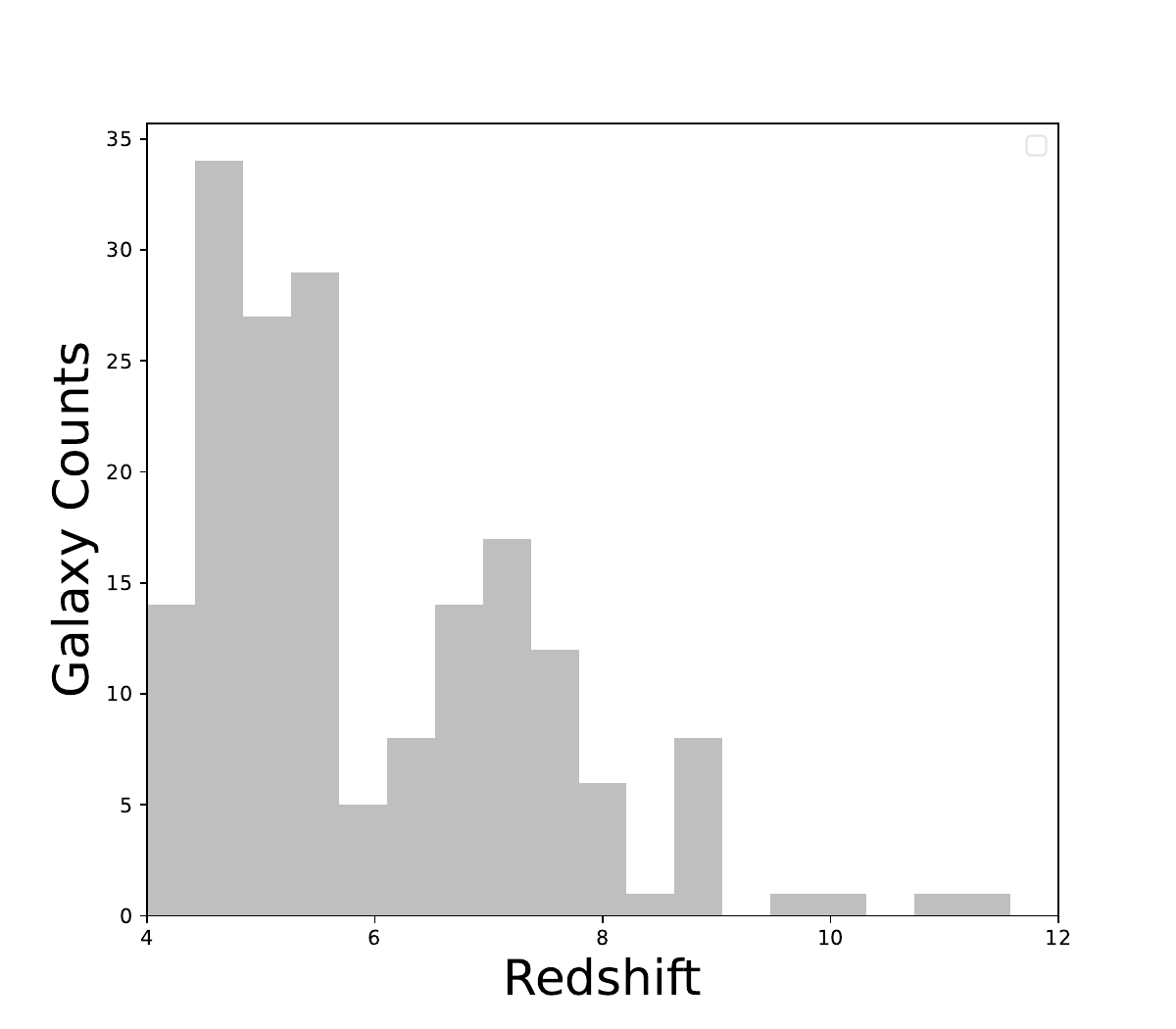}
    \caption{Distribution of spectroscopic redshifts derived by fitting the spectrophotometric data with CIGALE. We keep 173 objects with the most robust redshifts. Galaxies are mainly found in two subpopulations at 4 $\lesssim$ z $\lesssim$ 6 and 5$\lesssim$ z $\lesssim$ 6, with a tail extending to z $\lesssim$ 12.}
    \label{fig:redshifts}
    \end{centering}
\end{figure}

We have 1,337 spectroscopic observations with NIRSpec (\citealt{ArrabalHaro2023}) from CEERS. We use 634 of these NIRSpec observations carried out with the prism configuration. The spectroscopic targets were selected on the basis of CANDELS HST imaging (\citealt{Grogin2011, Koekemoer2011}) on various non-homogeneous criteria, which might introduce a bias in the selection. CEERS's NIRCam observations detected 101,808 objects photometrically (CEERS\_v0.51.4, \citealt{Bagley2023}). Whenever possible, we combine spectroscopic data (\citealt{Birkmann2022}) with photometric data by cross-matching the coordinates within 0.2 arcsec. However, because there is no complete overlap between CEERS NIRSpec and NIRCam observations, some NIRSpec fields are in areas where we have no NIRCam imaging. For them, we only use the spectroscopic data. After fitting the prism spectra (with and without NIRCam data), we check the quality of the spectroscopic redshifts for objects with z$_{spec}$ $>$ 4.0. We classify the redshifts in four classes of redshift quality from q$_z$=0 (no doubts on redshift), to q$_z$=4 (wrong or unconfirmed redshift). We keep 173 objects with NIRSpec observations, for which q$_z$=0 (all modeled lines match the observed spectrum) or q$_z$=1 (some fainter lines not in excellent agreement with the models). Objects with q$_z$=2 present a continuum that could be in agreement with the derived redshift, but without any positively identified lines, and objects with q$_z$=3 only have a low signal-to-noise hint for the estimated redshift. We have flagged a sample of 6 possible AGN based on the literature (\citealt{Kocevski2023, Larson2023, Harikane2023}). These AGN are listed in Tab.~\ref{Tab.AGNs}, and shown with crosses in the plots. From this analysis, the distribution of redshifts is shown in Fig.~\ref{fig:redshifts}. Most galaxies are clustered in the range z=4.0 to z=8.0, with a minority but important tail extending to z $\lesssim$12.

\begin{center}
\begin{table}
\caption{Extract from Harikane et al. (2023): possible AGNs in the analyzed sample.}
\begin{tabular}{|l|r|r|r|}
\hline
  \multicolumn{1}{|c|}{id CEERS} &
  \multicolumn{1}{c|}{redshift} &
  \multicolumn{1}{c|}{RA} &
  \multicolumn{1}{c|}{Dec} \\
\hline
nirspec4\_397    & 6.01 & 14:19:20.69 & +52:52:57.7\\
nirspec8\_717    & 6.94 & 14:20:19.54 & +52:58:19.9\\
nirspec4\_746    & 5.63 & 14:19:14.19 & +52:52:06.5\\
nirspec8\_1236   & 4.50 & 14:20:34.87 & +52:58:02.2\\
nirspec7\_1244   & 4.48 & 14:20:57.76 & +53:02:09.8\\
nirspec4\_2782   & 5.26 & 14:19:17.63 & +52:49:49.0\\
\hline
\end{tabular}
\tablefoot{Note that we keep these objects in the analysis but they are flagged in the figures.}
\tablebib{protect\cite{Harikane2023}}
\label{Tab.AGNs}
\end{table} 
\end{center}

\subsubsection{Analysis of spectrophotometric data}
\label{sect.data.analysis}

We use all data with a signal-to-noise ratio (SNR) SNR$>$1.0 per pixel in the spectrum and for each NIRCam photometric band. All other measures are set as upper limits in the fit. We utilize a new version of CIGALE that accepts both photometric and spectroscopic data (see the Appendix~\ref{appendix:CIGALE}) and we define this type of mixed data as spectrophotometric energy distributions (SPEDs) to clarify the difference from traditional spectral energy distributions (SEDs). The priors used in the fits are listed in Tab.~\ref{Tab.pcigale.ini} of the appendix \ref{sec:pcigale.ini} and a sample of the fits is shown in Figs.~\ref{fig:SEDs_1a} to \ref{fig:SEDs_2b}.

Two SFHs are used to test the stability of the results: a delayed-plus-burst and a periodic one. The delayed SFH assumes that star formation is active over a few tens to hundreds of Myrs with SFR(t) $\propto$ $\frac{t}{\tau^2} \times \exp^{-t/\tau}$ followed by a final burst with various possible ages. Various other types of SFHs could be used, but determining the SFH in the early universe is difficult for any SED modeling method (\citealt{Lower2020}). Moreover, \citealt{Leja2019, Iyer2019, Tacchella2023} insist on the fact that the priors chosen for the fit are the primary drivers, before the type of SFH, to recover the physical parameters. The number of priors thus sets strong constraints on the ability to successfully run fitting codes, especially for large samples. The speed of CIGALE (\citealt{Burgarella2024}) allows it to explore various sets of priors with several 10$^8$ models in a reasonable time for thousands of spectrophotometric objects. For the alternate SFH, a periodic SFH is chosen because it is conceptually different from the delayed-plus-burst SFH: it does not assume any kind of continuous SFH. Instead, a series of bursts, separated by regular quiescent periods, are used (see the appendix~\ref{appendix:periodic} that shows that this periodic SFH has a low impact on the results presented in this paper).

\begin{figure}
	\includegraphics[width=\columnwidth]{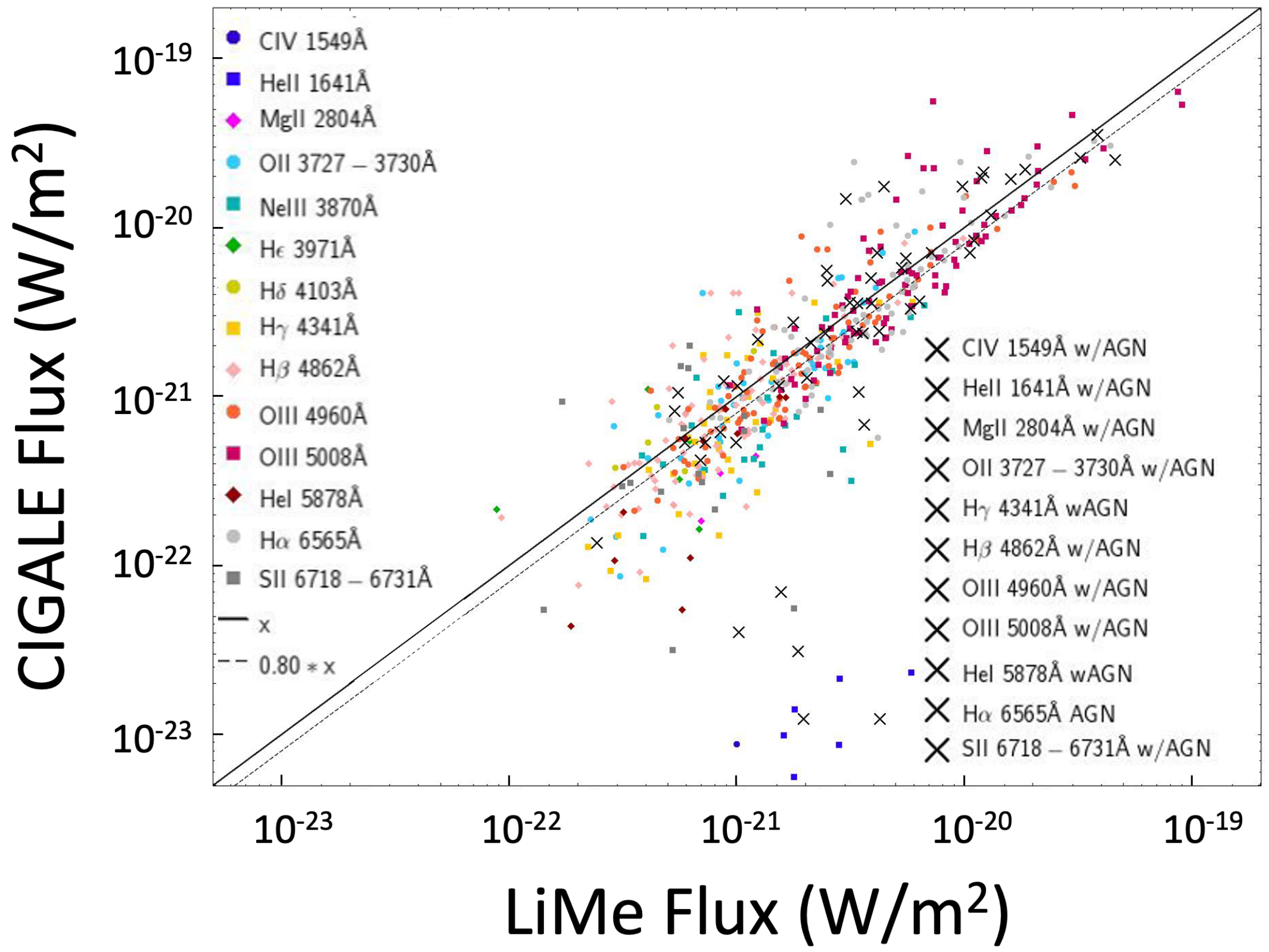}
    \caption{Comparison of the CIGALE fluxes to those measured with LiMe (\protect\citealt{Fernandez2024}). The full results on the CEERS catalog redshifts and line measurements will be discussed in Arrabal Haro et al. (in preparation). The points corresponding to each emission line are color-coded to better identify each species (see inside caption). AGN lines are identified by crosses. The most ultraviolet lines (CIV $\lambda$1549 $\r{A}$, HeII $\lambda$ 1641 $\r{A}$ in the lower part of the plot, are not in agreement with those estimated by LiMe as shown by the offset from the solid 1-to-1 line. Instead, we find that other lines are systematically different by about about 20 \% (dashed line). The observed difference might find an origin in the subtraction of the continuum.}
    \label{fig:fluxLime}
\end{figure}

\begin{figure}
	\includegraphics[width=\columnwidth]{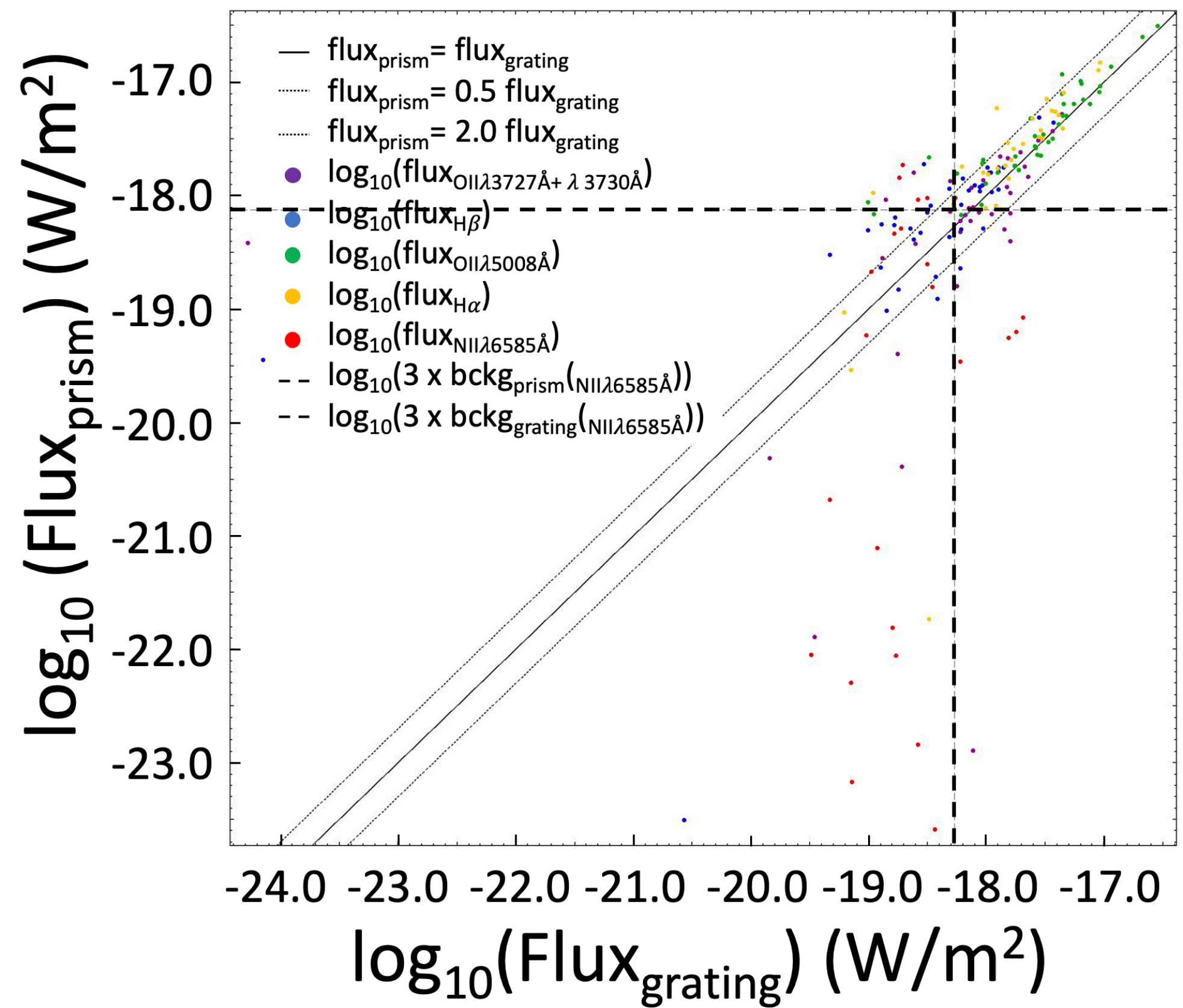}
    \caption{Comparison of the fluxes computed by CIGALE (Flux$_{prism}$) with Gaussian fitting of the lines observed in the grating configuration (Flux$_{grating}$), for objects observed in both configurations. The points corresponding to each emission line are color-coded to better identify each species (see inside caption). The diagonal line shows the one-to-one relation (Flux$_{prism}$ = Flux$_{grating}$) while the dashed lines present offsets (Flux$_{prism}$ = 0.5 Flux$_{grating}$ and (flux$_{prism}$ = 2.0 Flux$_{grating}$). The horizontal and vertical dashed lines show the level of the 3$\sigma$ background around the [NII]+H$\alpha$ lines.}
    \label{fig:fluxGaussian}
\end{figure}

CIGALE estimates line fluxes through a comparison with Cloudy-derived nebular models included in CIGALE. To be sure we are on a safe ground, we validate the line fluxes measured by CIGALE with those estimated via other methods: we check in Figs.~\ref{fig:fluxLime} and \ref{fig:fluxGaussian} that the flux of the emission lines measured by CIGALE are consistent with first fluxes estimated with the LiMe software (\protect\citealt{Fernandez2024}) and second, a fit performed on the sub-sample where both prism and grating NIRSpec spectra are available. For this second measurement, we fit the lines with lmfit (based on the Levenberg-Marquardt method). lmfit provides us with tools to perform non-linear optimization and curve fitting in Python. Practically speaking, we fit the lines per group of 3 lines (e.g. H$\beta$+[OIII] or [NII]+H$\alpha$) assuming a model which is the sum of 3 Gaussian distributions for the emission lines, plus a line to fit the continuum. The comparison is good up to line fluxes of about 3$\sigma$ of the local background.

\section{The properties of the galaxy sample}
\label{sec:DerivedProperties}

In this section, we define how we selected a specific population of 49 galaxies that have extremely low dust attenuation (GELDAs). We observationally define GELDAs using the following criteria: 
\begin{itemize}
    \item The FUV dust attenuation A$_{FUV}$ = 0.0 within 2$\sigma_{A\_FUV}$,
    \item The stellar mass M$_{star}$ $<$ 10$^9$M$_\odot$
\end{itemize}

\subsection{Star formation rate and stellar mass}
\label{sec:MainSequence}

\begin{figure}
	\includegraphics[width=0.8\columnwidth]{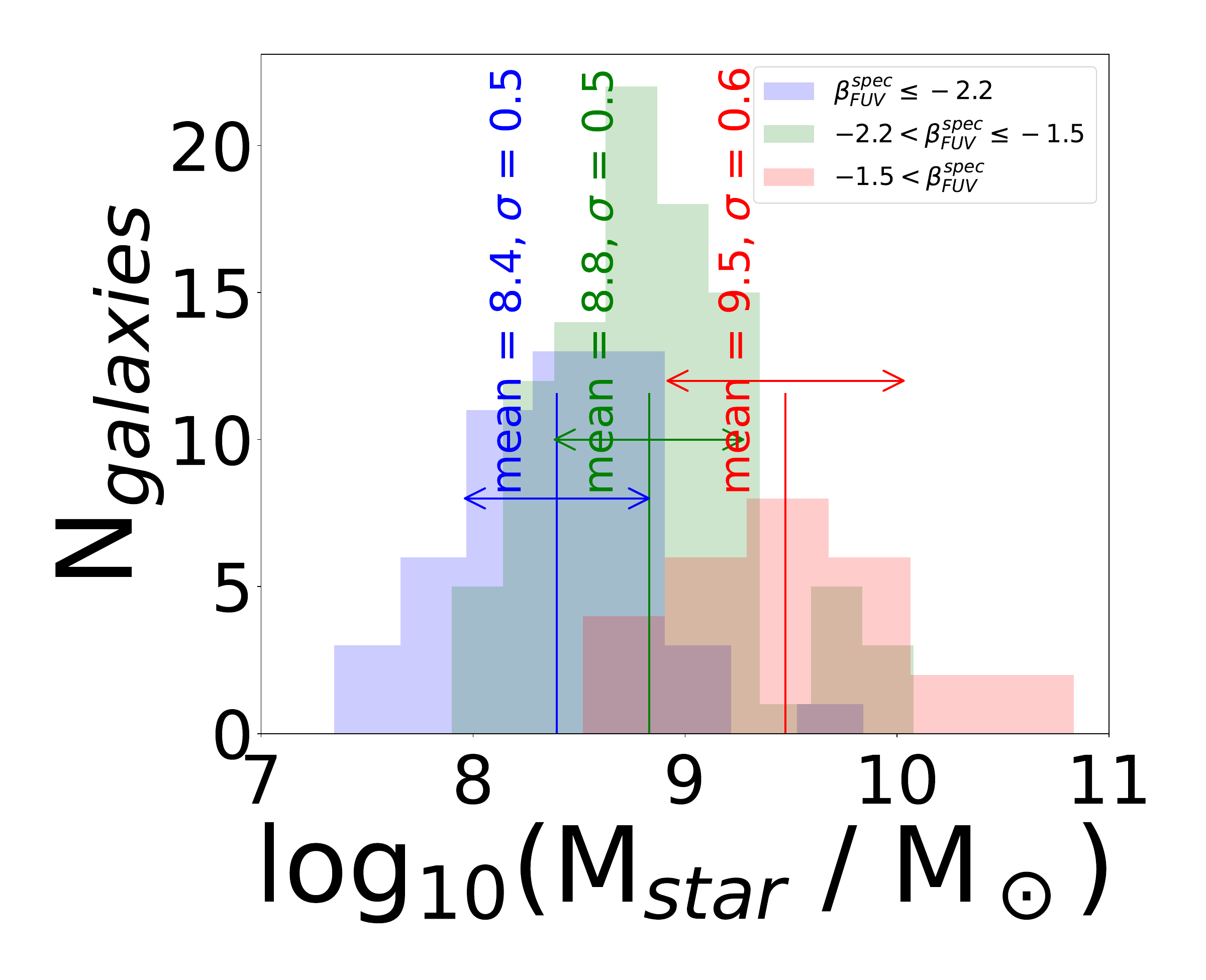}
    \caption{Distribution of stellar masses derived by fitting the spectrophotometric energy distributions of our galaxy sample. In this plot, we only show galaxies with $\beta_{FUV}$ derived from CIGALE's fits. The range of stellar masses reaches M$_{star}$ values as low as a few 10$^7$ M$_\odot$. The color coding shows that the least massive galaxies clearly have bluer UV slopes: $\beta_{FUV} \leq -2.2$ while the most massive galaxies have redder slopes: $-1.5 < \beta_{FUV}$, as expected from the dependence of UV dust attenuation on stellar mass (see, e.g. \citealt{Bogdanoska2020, Weibel2024, Bouwens2016}). The vertical numbers show the mean stellar mass for each sample.}
    \label{fig:histMstar}
\end{figure}

Fig.~\ref{fig:histMstar} presents the histogram of the stellar masses derived by CIGALE for the present sample of galaxies. The structure of the histogram as a function of the UV slope $\beta_{FUV}$ confirms the relation between stellar mass and dust attenuation at high and ultra-high redshifts as shown by several papers (\citealt{Bogdanoska2020, Weibel2024, Bouwens2016}). 

\begin{figure}
	\includegraphics[width=\columnwidth]{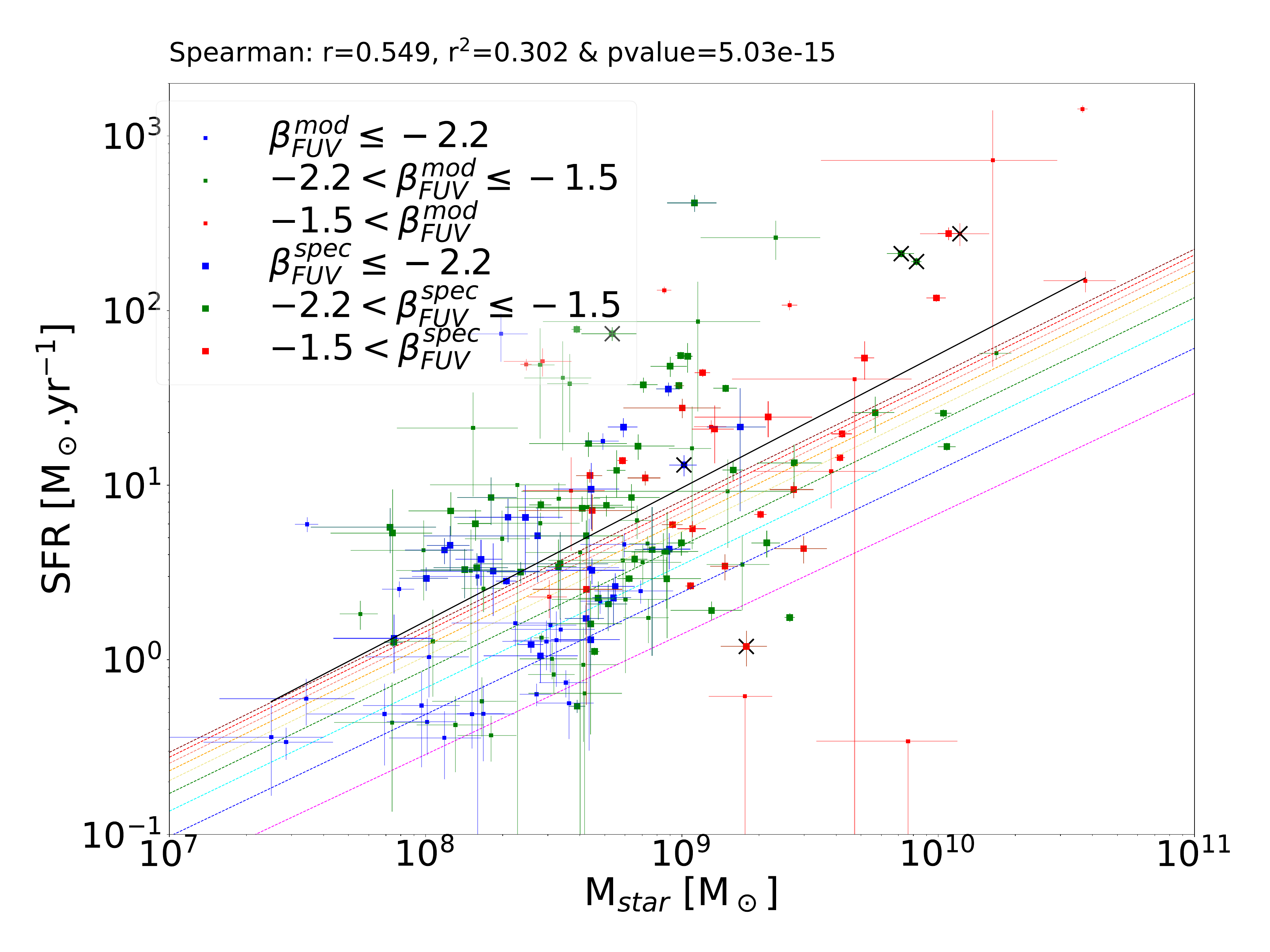}
    \caption{This plot presents the star formation rate as a function of the stellar mass for our galaxy sample. For some galaxies the observed spectra are not good enough to provide a trustworthy fit of the spectrum (f$_{\lambda} \propto \lambda^\beta_{FUV}$) for the UV slope. For those galaxies, we rely on CIGALE's fits. Here, the larger symbols correspond to UV slopes spectroscopically estimated from the observed data while the smaller symbols are from the CIGALE fits. Blue, green and red symbols respectively show galaxies with blue, intermediate and red slopes (see inside caption). Whatever the origin of the UV slopes, the lower mass galaxies present a bluer UV slope while the more massive have are redder. We superimpose in this plot the best-fit main sequence from \protect\cite{Speagle2014} in the range z = 4 to z = 12, color coded from magenta to dark red (bottom to top). }
    \label{fig:MainSequence}
\end{figure}

Fig.~\ref{fig:MainSequence} shows the location of the MS in the SFR vs. M$_{star}$ diagram. Because the redshift range is quite large, we might not expect all galaxies to follow a tight sequence. However, since most of the sample is in the redshift range 4$<$z$<$6 (Fig.~\ref{fig:redshifts}), we do observe such a sequence, which is structured by the UV slope $\beta_{FUV}$. We do not see any noticeable differences between UV slopes derived by directly fitting the spectra and those derived by fitting the SPEDs with CIGALE. In Fig.~\ref{fig:MainSequence}, we present a comparison of the location of our sample with the best-fit main sequence from \cite{Speagle2014}. The fits from \cite{Speagle2014} show the well-known strong increase in SFR at low redshifts (z$\lesssim$8) followed by a gradual lower increase at higher redshifts up to (z=12). Our sample of objects are mainly in the first redshift bin (4$\lesssim$z$\lesssim$8, see Fig.~\ref{fig:redshifts}). They are MS galaxies. However, part of the sample lies above the highest MS from \cite{Speagle2014}, especially the reddest. They likely belong to a star-busting class of galaxies. AGN are preferentially found at large masses and high SFRs.

\begin{figure}
	\includegraphics[width=\columnwidth]{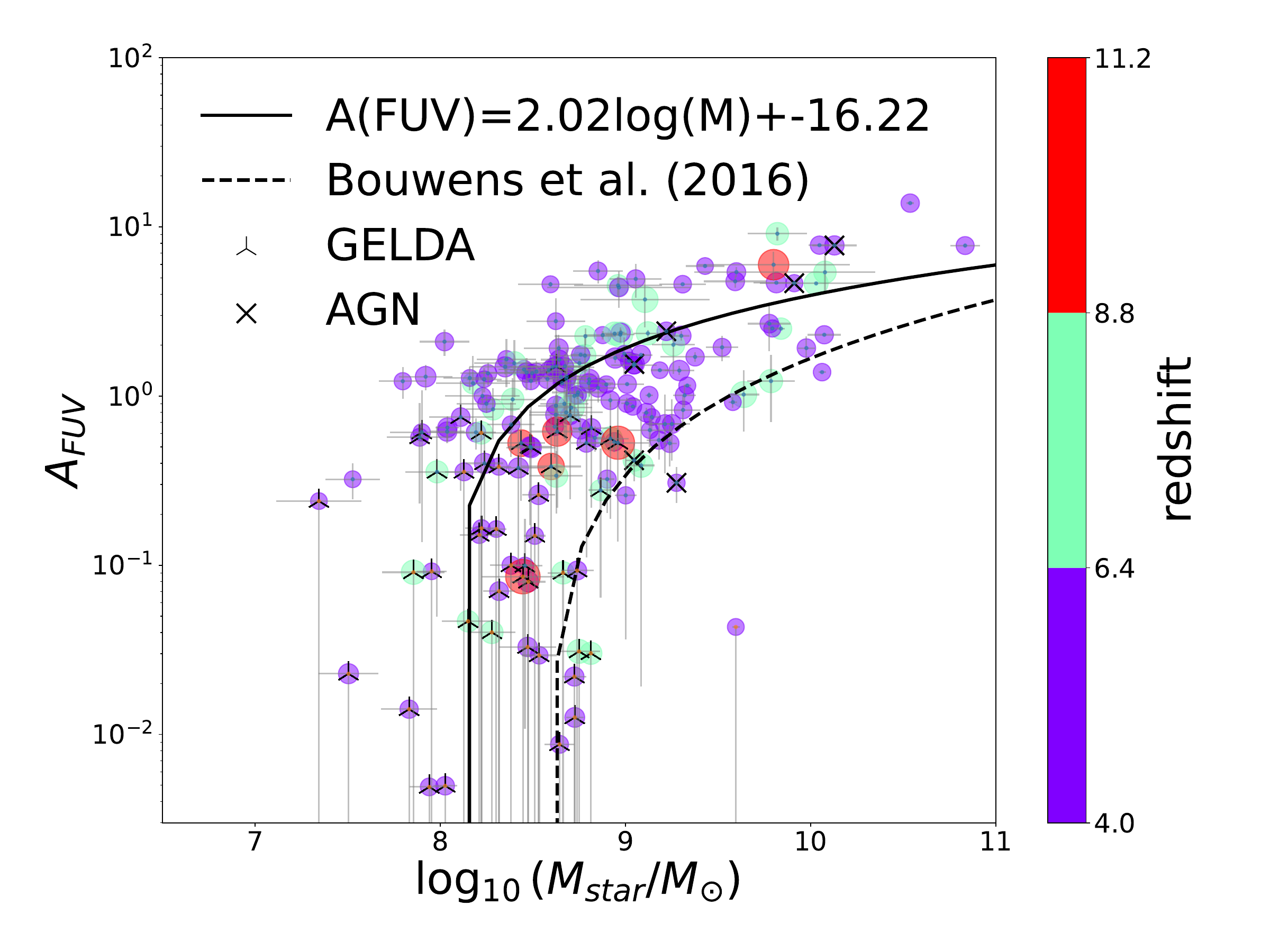}
    \caption{Far-UV dust attenuation A$_{FUV}$ as a function of the stellar mass, $\log_{10}$(M$_{star})$. The symbols are color-coded and sized with the redshift. The red objects corresponds to the highest redshift bin defined in this paper (z $>$ 8.8). The majority of highest-redshift GELDAs (5 out of 6) are found in the low-A$_{FUV}$ part of the plot. This sample of galaxies at z $>$ 8.8 is small, but it might confirm an expected decreasing evolution of the dust attenuation with the redshift. The dashed line is the ‘consensus’ law from \protect\cite{Bouwens2016} while the continuous line is the best linear fit (A$_{FUV}$ vs $\log_{10}$(M$_{star}$)) to the present data. GELDAs are shown with upside-down Y and AGN with crosses.}
    \label{fig:AFUV-Mstar}
\end{figure}

In Fig.~\ref{fig:AFUV-Mstar}, we show the far-UV dust attenuations, A$_{FUV}$, vs. $\log_{10}$(M$_{star}$) where we again note a large dispersion at low M$_{star}$ spanning more than 2 decades. The stellar mass is not the only acting parameter as both low and relatively high A$_{FUV}$ lie in the same stellar mass range. We do not observe any clustering of low-redshift galaxies (z $\lesssim$ 8.8) in the lower part of the plot. This shows that these low-redshift galaxies could have a wide range of dust attenuation. On the other hand, five out of six of the highest redshift galaxies (z $\gtrsim$ 8.8) are found in this part of Fig.~\ref{fig:AFUV-Mstar}, and are therefore GELDAs. This would suggest that GELDAs become dominant in the early Universe.

The ‘consensus’ law between A$_{FUV}$ and M$_{star}$ estimated at the lower redshift (z$\sim$2-3) by \cite{Bouwens2016} does not pass through the present data. \cite{Bogdanoska2020} found that this ‘consensus’ law does not appear to be valid at large redshifts. They also suggested that the low stellar mass galaxies exhibit a large scatter in A$_{FUV}$ and proposed an evolution of this A$_{FUV}$ and M$_{star}$ relation with the redshift. However, the sample available in \cite{Bogdanoska2020} could hardly reach $\log_{10}$(M$_{star}$) $\lesssim$ 9.0. JWST sensitivity to much fainter flux allows one to reach galaxies at much lower stellar masses, and potentially also permits one to detect this new population of GELDAs.

\subsection{The mass of metals and gas of the galaxy sample}
\label{sec:metal-gas-mass}

In CIGALE, the stellar Z$_{star}$ and nebular Z$_{gas}$ metallicities have different priors. This is most useful when spectroscopic data are available because they allow to separately constrained both metallicities by fitting the spectra, including lines. For this specific estimation process, CIGALE makes use of the new nebular models described in \cite{Theule2024} that can have excitation parameters up to logU = -1.0 and with a wide range of nebular metallicities and electronic densities (n$_H$). The metallicity Z$_{gas}$ derived by CIGALE can be converted into 12+$\log_{10}$(O/H) where O/H is the oxygen abundance of the gas by using their table 1 (correspondence between $\xi_0$, the interstellar gas metallicities and the stellar metallicities). $\xi_0$ is defined on oxygen abundance: $\xi_0$ = (O/H) / (O/H)$_{GC}$, where (O/H)$_{GC}$ = 5.76 $\times$ 10$^{-4}$, and GC is the so-called local Galactic concordance (\cite{Theule2024} follows \citealt{Nicholls2017}). From this, Eq.~1 links total metallicity to oxygen abundance:

$${\rm Eq.~1:~} 12+\log_{10}(O/H)) = \log_{10}(Z_{gas}) + 10.410.$$

\begin{figure}
    \begin{centering}
	\includegraphics[width=0.8\columnwidth]{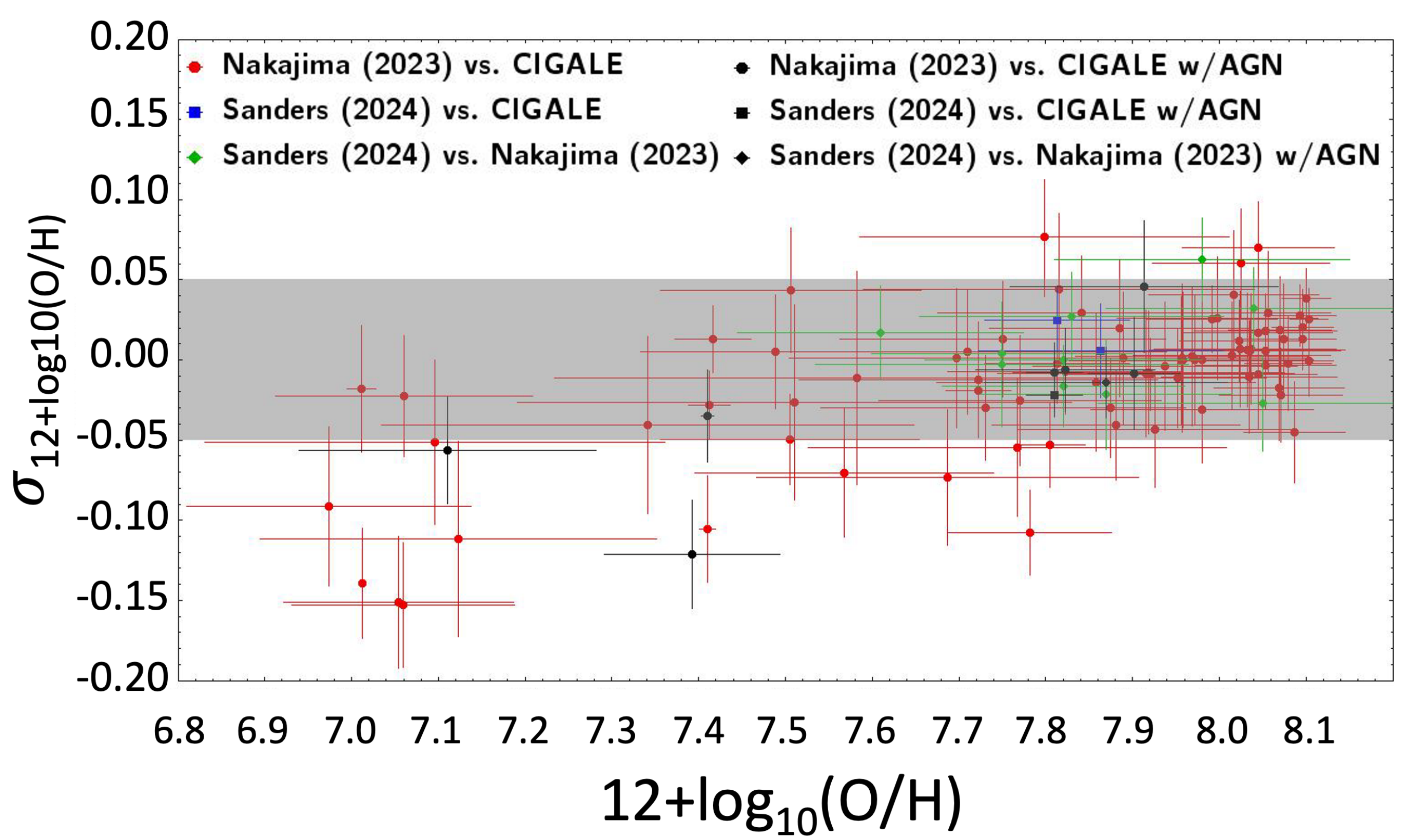}
    \caption{Difference between metallicity estimates (12+$\log_{10}(O/H)$ defined in Eq. 1): a comparison of the metallicities measured for some of our galaxies in the literature (\protect\citealt{Nakajima2022, Nakajima2023, Sanders2024}) via auroral lines, with our own estimates. At relatively high metallicities, above 12+log10(O/H)$\sim$7.4-7.6 (that is Z/Z$_{gas}$~0.1-11\% Z$_\odot$), the differences remain within $\sigma$(12+$\log_{10}(OH))$$<$0.05. This is about the same dispersion between other metallicity estimates, although on a smaller sample. However, there is a disagreement at lower metallicities, and CIGALE's 12+$\log_{10}$(O/H) present an offset that might be systematic or increasing with lower metallicities by about $\sigma$ ($\log_{10}$(O/H))$<$0.10. But the direct method is not well calibrated at $\log_{10}$(O/H)$<$7.4-7.6 because there are less than 5 objects with auroral lines at such low metallicities estimated via the direct-method oxygen abundances. }
    \label{fig:metallicity1}
    \end{centering}
\end{figure}

In Fig.~\ref{fig:metallicity1}, we compare the CIGALE metallicities estimated for the same CEERS galaxies by \cite{Nakajima2022, Nakajima2023} and \cite{Sanders2024}. \cite{Nakajima2023} measured emission line fluxes for 135 galaxies (115 in CEERS). For 10 of these galaxies, they determined their electron temperatures with [O III] $\lambda$4363 $\r{A}$ lines, in a way similar to lower redshift star-forming galaxies, and they derived the metallicities by a direct method. They finally estimated metallicities for their entire sample of JWST-observed galaxies with strong lines using their previous metallicity calibration (\citealt{Nakajima2022}), based on the direct method measurements. Our sample common with \cite{Nakajima2023} amounts to 90 galaxies. \cite{Sanders2024} also combine JWST measurements with [O III] $\lambda$4363 $\r{A}$ auroral line detections from JWST/NIRSpec and from ground-based spectroscopy to derive electron temperature (T$_e$) and direct-method oxygen abundances on a combined sample of 12 star-forming galaxies at z=1.4-8.7. Our sample shares 3 of them, for which we derive metallicities with CIGALE. Finally, \cite{Nakajima2023} and \cite{Sanders2024} have 3 common objects. Our fitting method is different, as the total spectrophotometric fits allow one to consistently constrain the metallicities (Z$_{star}$ and Z$_{gas}$) by selecting only models that agree with the whole information brought by observations, that is, continuum and lines, together. 

\begin{figure}
	\includegraphics[width=\columnwidth]{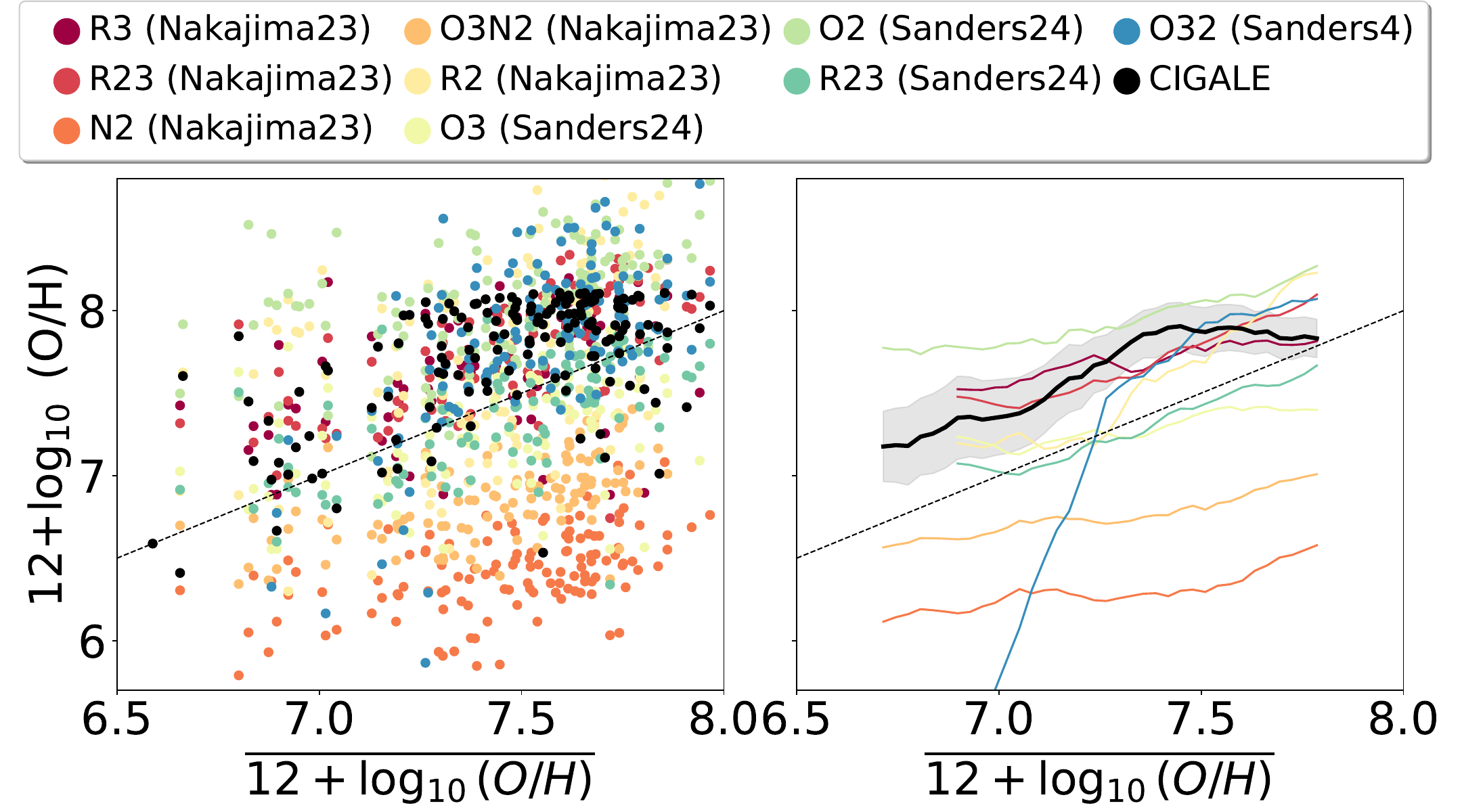}
    \caption{Direct comparison of metallicities (12+$\log_{10}(O/H)$ defined in Eq. 1. Left: this panel shows the metallicities derived using various published calibrations as a function of the mean metallicities computed from each measured galaxy. The dashed black line is the 1-to-1 relation. For a given metallicity on an x-axis, we could get a very wide range on the y-axis depending on which line ratio is used. The black dots are the metallicities from CIGALE. The red-orange ones from \cite{Nakajima2023} and the blue-green one from \cite{Sanders2024}. The color coding is given in the legend inside the plot. 
    Right: this panel presents the same information with a rolling average. Most of the metallicity values agree within about $\pm$ 0.5 dex and the metallicities from CIGALE are approximately in the middle range. However, we notice that the ones involving the nitrogen lines lead to much lower metallicities (see text for more details.}
    \label{fig:metallicity2}
\end{figure}

The metallicities estimated by CIGALE show a systematic or gradually increasing discrepancy at $\log_{10}(O/H)<7.4-7.6$ with \cite{Nakajima2023} with metallicities lower by about $\sigma$ ($\log_{10}$(O/H))<0.10. However, we note that the direct method is not well calibrated below $\log_{10}(O/H)<7.4-7.6$ with less auroral measurements for objects at these low metallicities. The true value becomes quite uncertain until we could get a better calibration. Fig.~\ref{fig:metallicity2} shows that individually and globally the metallicity estimates can vary depending on the method used to derive metallicities from the lines. The CIGALE metallicities estimated by fitting the entire spectrum are found close to most values from \cite{Nakajima2023} and \cite{Sanders2024}, and especially very close to the R23 index which is found to be the most reliable among various metallicity indicators over the wider range of metallicity (\cite{Nakajima2022}). However, the line ratios involving nitrogen lines (N2 and O3N2 from \cite{Nakajima2023} lead to much lower metallicities by about 1 dex. \cite{Nakajima2022} found a scatter as large as $\Delta\log_{10}$(O/H) $\sim$ 0.4 dex in the relation for metallicities derived using the N2 index. This is especially true at low metallicities, as for our galaxies. They suggested that this might be associated with line ratios that use single-ionized, low-ionization lines such as [NII]. We conclude that care should be taken when using only one of these indices. However, CIGALE provides us with a safe and reliable method for estimating metallicities, at least in the present range.

\begin{figure}
    \includegraphics[width=0.72\columnwidth, angle=-90]{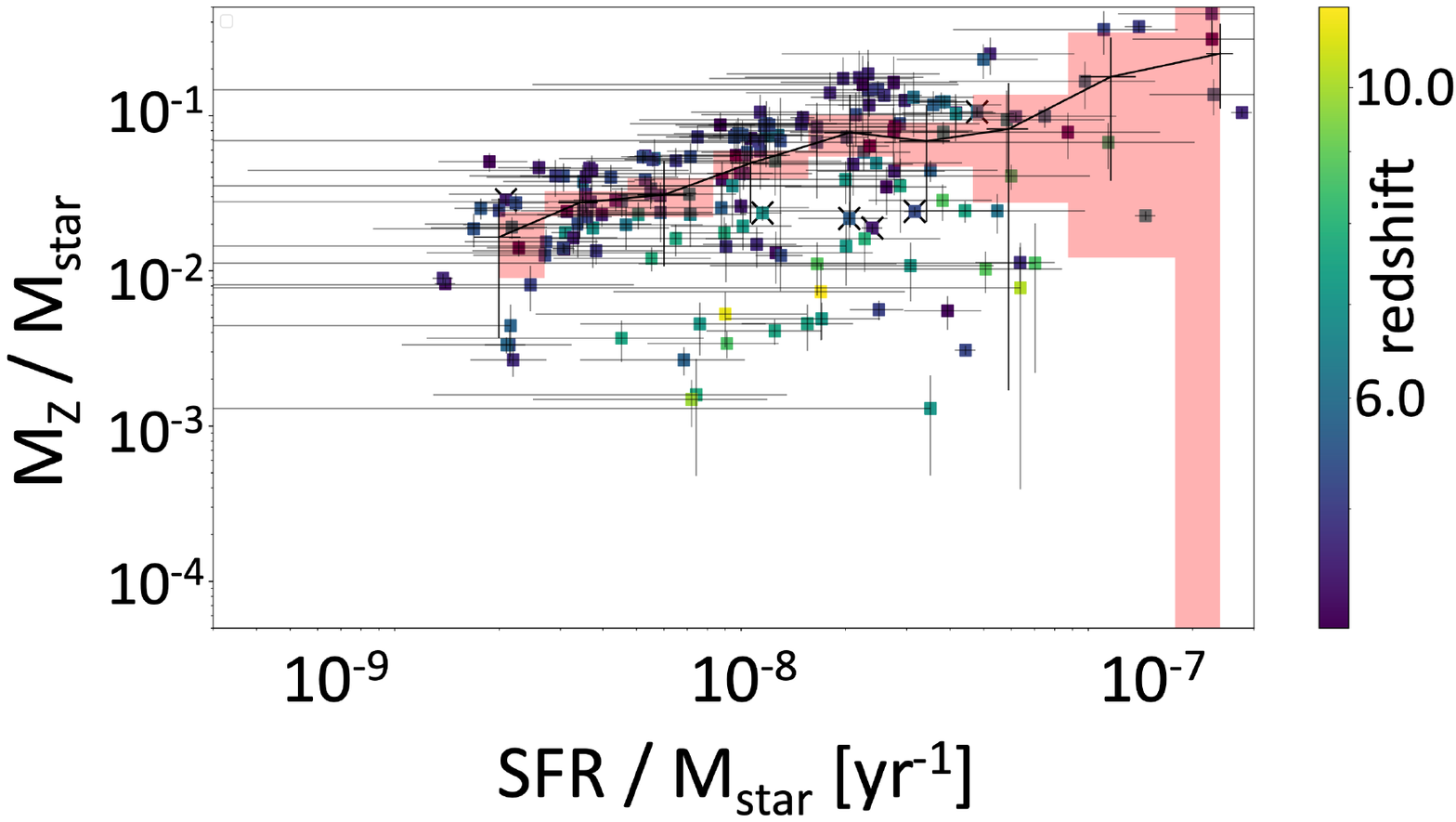}
    \caption{We present the evolution of the metal-to-stellar mass ratio as a function of the specific star formation rate in this metal formation rate diagram (MFRD) in the same way we defined the dust formation rate diagram (DFRD) that shows how the specific dust mass (M$_{dust}$/M$_{star}$) changes with the specific star formation rate (SFR/M$_{star}$) in \protect\cite{Burgarella2022}. This type of diagram is normalized to M$_{star}$ and allows a better comparison between galaxies. The symbols are color-coded with the redshift. The pink-shaded area presents the 2-$\sigma$ confidence interval for the distribution in $M_Z$/M$_{star}$, and the black solid line shows the rolling average. We observe a decreasing trend with decreasing sSFR. The highest-redshift galaxies are located in the lowest part of the plot.}
    \label{fig:MZMstar-sSFR}
\end{figure}

Fig.~\ref{fig:MZMstar-sSFR} presents the specific mass of metals (M$_Z$/M$_{star}$) for the present sample as a function of the specific SFR (SFR/M$_{star}$), where the metal mass, M$_Z$ is computed from Eq. 2 (\citealt{Heintz2023a}):

$${\rm Eq.~2:~} M_Z = M_{gas} \times 10^{12+log_{10} (O/H)-8.69} \times  Z_\odot$$

In Eq. 2, the gas mass, M$_{gas}$, and the oxygen abundance, 12+$\log_{10}$(O/H), are estimated via the spectrophotometric fitting and from Eq. 1. To estimate M$_{gas}$, we need to estimate the molecular mass M$_{molgas}$ from Eq. 3 (\citealt{Tacconi2020}). 

$${\rm Eq.~3:~} log_{10} (M_{molgas}) = $$
$$0.06 - 3.33 \times [log_{10}(1 + z) - 0.65]^2 + $$$$0.51 \times log_{10}(sSFR/sSFR(MS, z, M_{star}) $$
$$- 0.41 \times [log_{10}(M_{star}) - 10.7] \times M_{star}$$

where the specific instantaneous SFR, sSFR=SFR$_{inst}$/M$_{star}$, is derived from the spectrophotometric fitting, while the reference sSFR for the MS as a function of the stellar mass M$_{star}$, and the redshift: sSFR(MS, z, M$_{star}$) is from \cite{Speagle2014}, assuming the so-called "Bluer” w/ high-z obs" MS in their Table 9 (Eq. 4). The MS is modeled up to z$\sim$6, while our sample reaches z=11.4. However, even if the scatter in the MS is quite large, studies suggest that it should not show any strong evolution to z$\sim$12 (\citealt{Cole2025, Chakraborty2024}). 

$${\rm Eq.~4:~} \log_{10} (SFR (M_{star}, Age_{Universe})) = $$$$[(0.73 - 0.027 \times Age_{Universe} ) \times \log_{10} (M_{star}) - $$$$(5.42 + 0.42  \times Age_{Universe}) ] - \log_{10} (M_{star})$$

We also need to add the contribution from the atomic gas M$_{atomgas}$ to M$_{molgas}$ The atomic-to-molecular mass ratio M$_{atomgas}$/M$_{molgas}$ is estimated by \cite{Chowdhury2022} for star-forming galaxies at z$\sim$0, z$\sim$1.0, and z$\sim$1.3 with values in the range 4$\pm$2 for galaxies with M$_{star}$$>$10$^{10}$ M$_\odot$. To account for galaxies with $<$10$^{10}$ M$_\odot$ in their statistics, they assume that the ratios M$_{atomgas}$ to M$_{molgas}$ are systematically higher by a factor of about 5 for all galaxies with M$_{star}$ $<$ 10$^{10}$M$_\odot$. In this case, the values obtained would increase M$_{molgas}$ at z $\sim$ 1.3 by a factor of approximately 2, giving a ratio M$_{atomgas}$/M$_{molgas}$ = 2.5 for the highest redshifts. At higher redshifts (0.01 $<$ z $<$ 6.4), there is no significant redshift evolution of the M$_{atomgas}$/M$_{molgas}$ ratio (\citealt{Messias2024}), which is about 1-3. At z=8.496, the gas and stellar contents of a metal-poor galaxy are studied with JWST and ALMA (\citealt{Heintz2023b}). From this analysis, they infer M$_{molgas}$ = (3.0-5.0) $\times$ 10$^8$ M$_\odot$. corresponding to 40\% $\pm$ 10\% of M$_{gas}$ for their object, which leads to M$_{atomgas}$/M$_{molgas}$ = 1.5$^{2.0}_{1.3}$. Given the redshift of our objects, we will assume in this paper M$_{atomgas}$/M$_{molgas}$=2.0$^{2.5}_{1.3}$.

\subsection{Dust masses}
\label{sec:Mdust}

Dust masses (M$_{dust}$) are very important for building diagnostics on the origin of dust in galaxies because they are directly related to the dust building rate, which in turn can provide us with information on the origin of dust grains (e.g., \citealt{Lesniewska2019, Nanni2020, Burgarella2022}). To help estimate the dust masses for our objects, we use deep 450 and 850 $\mu$m SCUBA-2 and NOEMA-1.1 mm observations. Both are cross-correlated with JWST’s coordinates. The former have a mean depth of $\sigma_{450}$=1.9 and $\sigma_{480}$=0.46 mJy beam$^{-1}$ and the angular resolution is $\theta_{FWHM}$ $\approx$ 8 arcsec at 450 $\mu$m, and $\theta_{FWHM}$ $\approx$ 14.5 arcsec at 850 $\mu$m. For NOEMA, the rms is $\sigma_{1.1 mm}$=0.10 mJy beam$^{-1}$, and the beam size is 1.35 $\times$ 0.85 arcsec$^2$. Although the sensitivity of these observations is typically lower than that required to detect most of our galaxies, the inferred upper limits are useful to put constraints on the total IR luminosities and dust masses. Similarly, the angular resolutions are much larger than JWST’s (\citealt{Zavala2023}). Some of the associations might therefore be wrong. However, these sub-mm data rule out any strong lower-redshift far-IR emitters that would be associated with the objects in our sample.

The far-IR information for this sample is limited and we cannot directly derive any information on the dust emission SED shape. However, the ALMA-ALPINE sample {e.g. \citealt{Bethermin2020, Pozzi2021, Sommovigo2022a} and the ALMA-REBELS samples (e.g. \citealt{Inami2022, Sommovigo2022b, Algera2024a} and \citealt{Algera2024b}} presents physical properties, e.g. stellar masses: $\log_{10}$(M$_{star}$)$\sim$10, redshifts: 4.5$<$z$<$7.7 similar to ours (\citealt{Burgarella2022, Nanni2020}). Due to the similarities of the samples, we will assume that we can make use of the same \cite{Draine2014} best-fit model (see Tab~\ref{Tab.pcigale.ini}) identified in \cite{Burgarella2022, Nanni2020}. We note that this ALMA-ALPINE model corresponds to a dust temperature T$_{dust}$=54.1$\pm$6.7 K, assuming an optically thin modified black-body, which is in good agreement with \cite{Sommovigo2022a} who found an average value $<T_{dust}>$=48 $\pm$ 8 K and M$_{dust}$ in the range (0.5-25.1)$\times$ 10$^7$ M$_{\odot}$ for ALPINE. For ALMA-REBELS, the median T$_{dust}$ is in the 39-58 K range, and the median dust masses are estimated in the range (0.9-3.6)$\times$ 10$^7$ M$_{\odot}$ (\citealt{Sommovigo2022b}). \cite{Sommovigo2022b} also predict that dust masses can be produced by SNae alone for 85 \% of the REBELS sample. We note that \cite{Algera2024a, Algera2024b} found lower dust temperatures (T$_{dust}$=30 - 35 K) for two objects in the REBELS sample. Furthermore, \cite{Sommovigo2022a} predict that more metal-poor high-z galaxies could have warmer temperatures because of their smaller dust content, while the objectives studied in \cite{Algera2024a, Algera2024b} are metal-rich. However, the dust model is assumed to be the same in our work for all of our objects. Globally modifying our model would cause an offset of all the dust masses, but not the observed relative difference between GELDAs and non-GELDAs.

\begin{figure}
	\includegraphics[width=\columnwidth]{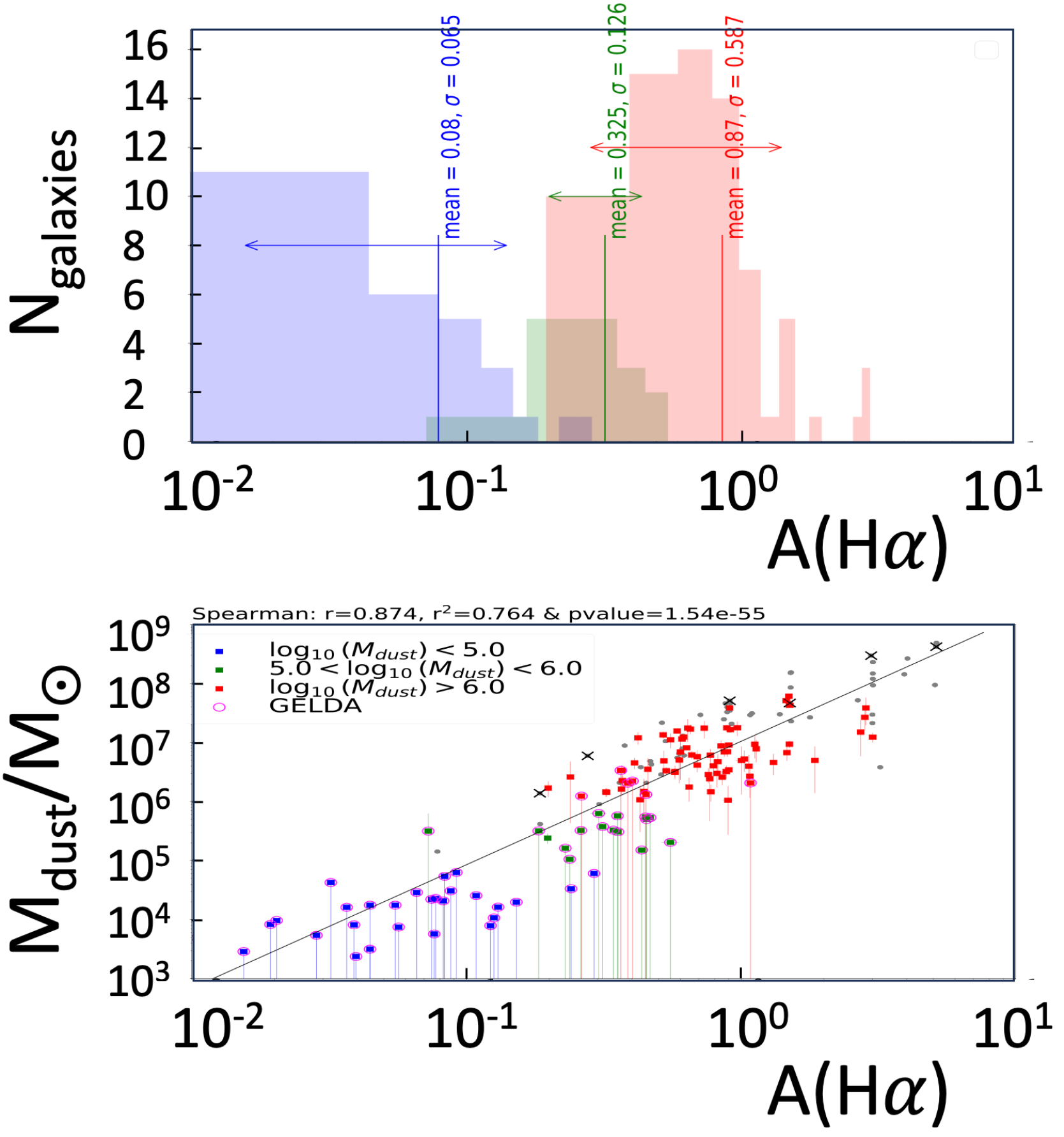}
    \caption{Correlation of the estimated dust mass, M$_{dust}$ with the dust attenuation, A(H$\alpha$). The most relevant parameter to predict the dust mass is the dust attenuation A(H$\alpha$) derived from the H$\alpha$/H$\beta$ Balmer decrement. The correlation with A(H$\alpha$) alone accounts for 76\% of the variation in M$_{dust}$. The spectral information brought by NIRSpec is thus fundamental to estimate the dust masses. Blue, green and red symbols respectively mean M$_{dust}$ $\leq$ 10$^5$ M$_\odot$, 10$^5$ $<$ M$_{dust}$ $\leq$ 10$^6$ M$_\odot$ and M$_{dust}$ $>$ 10$^6$, while magenta circles show GELDAs.}
    \label{fig:correlations_AHalpha}
\end{figure}

\begin{figure}
	\includegraphics[width=\columnwidth]{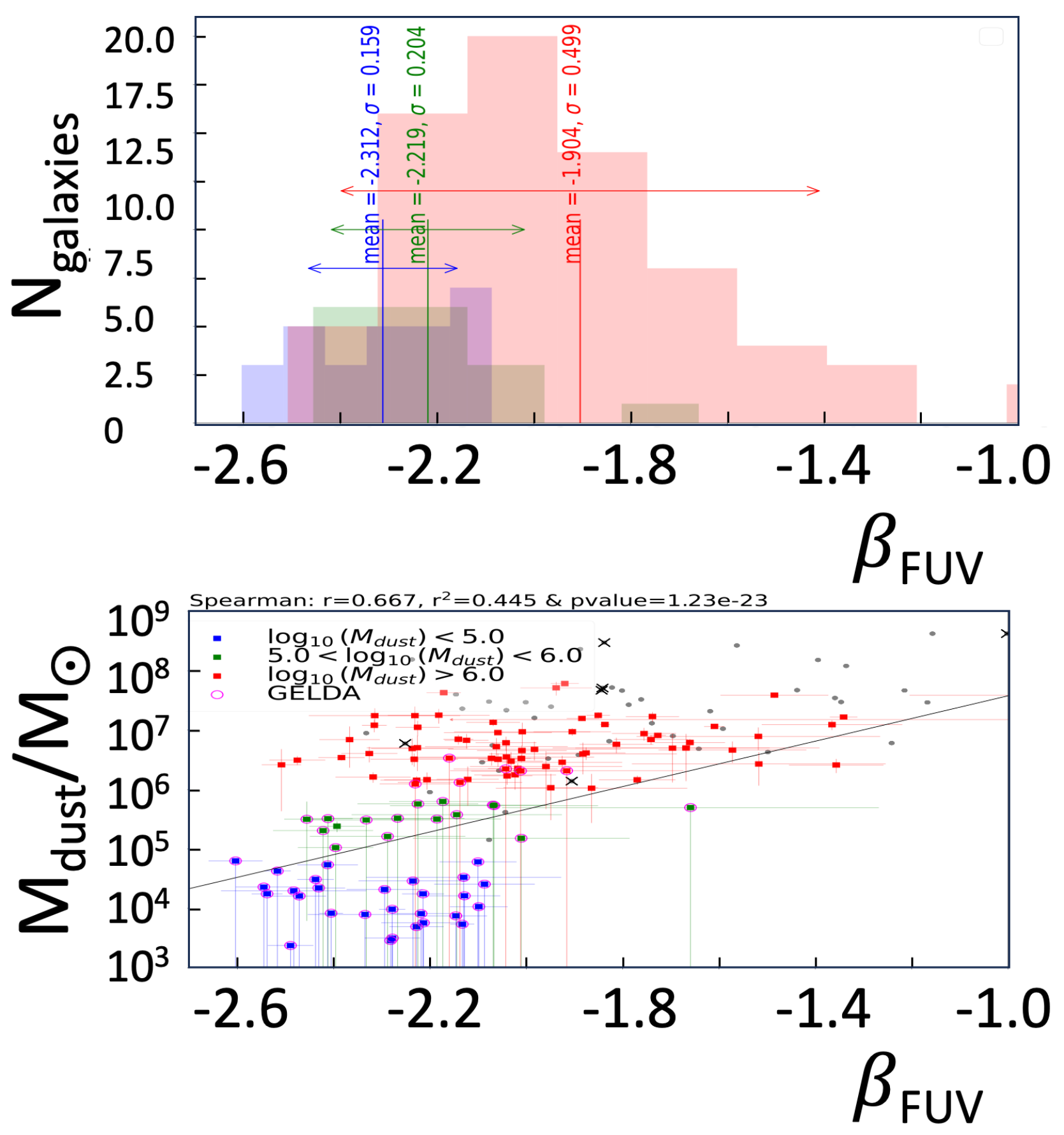}
    \caption{Same colors as in Fig.~\ref{fig:correlations_AHalpha}. Correlation of the estimated dust mass, M$_{dust}$ with the UV slope $\beta_{FUV}$. The correlation with $\beta_{FUV}$ alone accounts for 44 \% of this variation, which confirms that the spectral information on lines is the most important one. The other tested parameters: metallicity (about 2 \%) and redshift ($<$ 1\%), but also the level of the sub-mm upper limits are not significantly correlated with M$_{dust}$ in this analysis.}
    
    \label{fig:correlations_betaFUV}
\end{figure}

Because we use the above single dust emission model (\citealt{Draine2014}), we do not derive any shape for the IR emission in this paper. The shape of the IR spectrum is fixed by the ALMA-ALPINE sample at 4.5$<$z$<$6.2, and the IR luminosity is estimated assuming the energy balance concept; M$_{dust}$ is constrained by the amount of dust attenuation and by the main observables that define this dust attenuation. Information on the amount of dust attenuation comes from the line ratios, especially H$\alpha$/H$\beta$ when available, the UV slope $\beta_{FUV}$, and from any available IR/sub-mm data (Figs.~\ref{fig:correlations_AHalpha} and \ref{fig:correlations_betaFUV}).

The analysis of the results suggests that the SCUBA-2 sub-mm fluxes do not significantly help in constraining the dust mass (see Fig.\ref{fig:MdustMstar-others}), as we do not find any correlation between the measured fluxes or upper limits. NOEMA detections are deeper and provide flux densities that are more useful in constraining the dust mass. However, we only have two objects in the sample and none of them within the lower sequence of GELDAs.
The Balmer decrement H$\alpha$/H$\beta$ and the dust attenuation for H$\alpha$, A(H$\alpha$) are strongly correlated with M$_{dust}$ (see Fig.~\ref{fig:correlations_AHalpha}) correlation coefficient r$_{H\alpha/H\beta}$=0.874). The UV slope $\beta_{FUV}$ is also (see Fig.~\ref{fig:correlations_betaFUV}), although at a lower level, correlated with M$_{dust}$.(correlation coefficient r$_{\beta\_FUV}$=0.667). We can thus conclude that, first, the emission lines, and second, the continuum shape drive the estimation of the amount of energy transferred into the far-IR. For our galaxy sample, the observed NIRSpec spectrum from about 0.5 to 5.3 $\mu$m provides the best spectral information to estimate the amount of dust attenuation via the Balmer decrement and the UV slope. Then, using the energy balance hypothesis, we can estimate the IR luminosity and thus M$_{dust}$, if the IR spectrum from \cite{Burgarella2022} is assumed to be valid for our present sample.

\begin{figure}
    \begin{centering}
	\includegraphics[width=0.8\columnwidth]{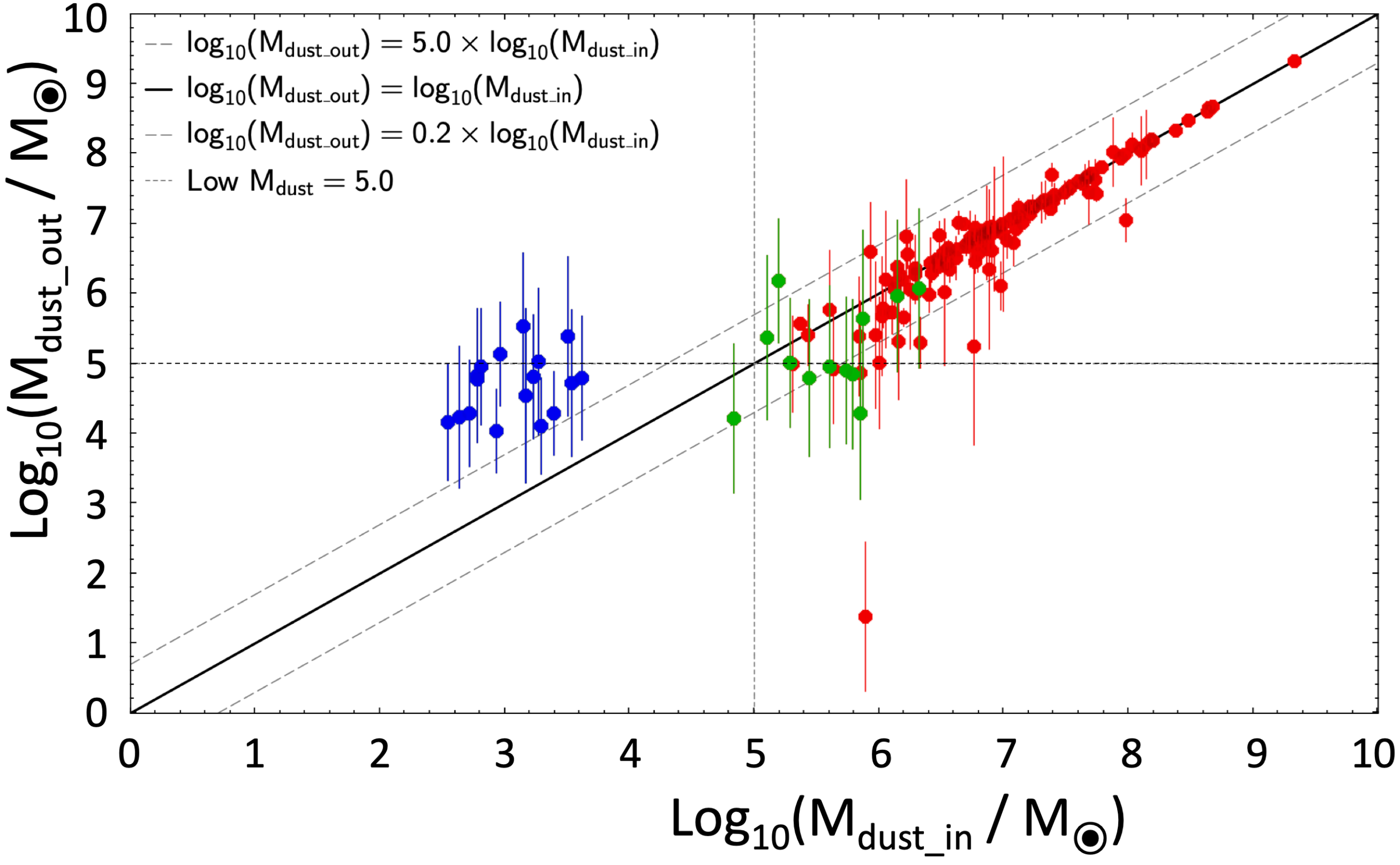}
	\includegraphics[angle=-90, width=0.8\columnwidth]{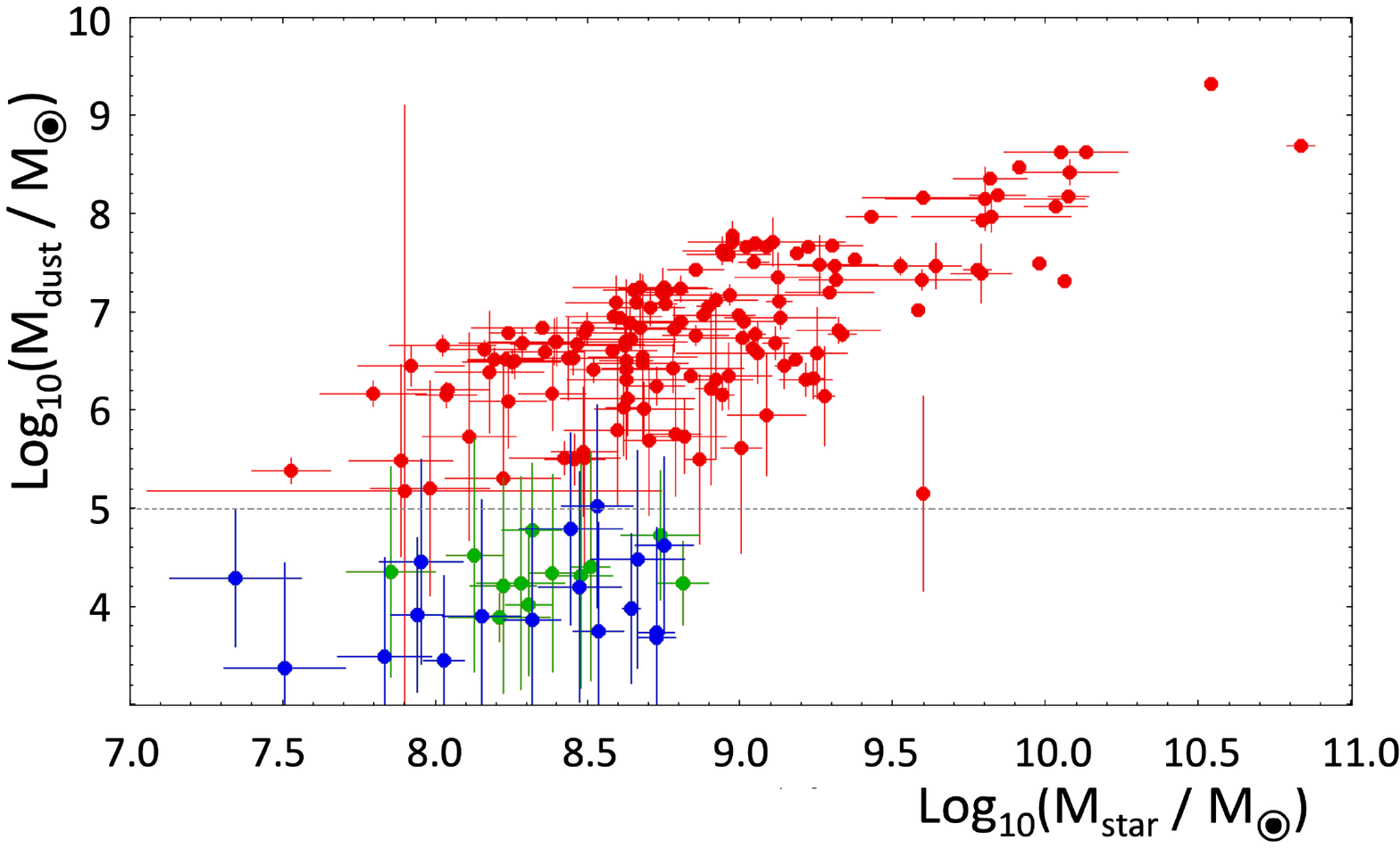}
    \caption{Use of a mock analysis to estimate the minimum dust mass that we can actually estimate using our approach. Top: The x-axis shows the modeled M$_{dust}$ computed from the best-fit models and for each of our galaxies. CIGALE is able to recover (y-axis) by fitting the mock data, these input dust masses, down to about $\log_{10}$(M$_{dust}$) = 5.0. Parts of the objects in green and all the objects in blue should thus be considered as upper limits. These objects that are the ones identified as transitioners between the upper sequence and a possible lower sequence, are well below the prime sequence (red dots). Bottom: in this $\log_{10}$(M$_{dust})$ vs. $\log_{10}$(M$_{star})$ diagram, we can see that the lowest dust masses correspond to the lowest stellar masses. However, for the same stellar mass range, we do observe a wide range of dust masses. Normal, mid and low M$_{dust}$ respectively mean M$_{dust}$ $\leq$ 10$^5$ M$_\odot$, 10$^5$ $<$ M$_{dust}$ $\leq$ 10$^6$ M$_\odot$ and M$_{dust}$ $>$ 10$^6$  M$_\odot$.}
    \label{fig:limit_Mdust}
    \end{centering}
\end{figure}

We perform tests that suggest that the minimum dust mass that we could estimate with CIGALE with the above assumptions is $\log_{10}$M$_{dust}$ = 5.0 (Fig.~\ref{fig:limit_Mdust}). To perform these tests, we use the ability of CIGALE to create a mock catalog based on the best-fit SPEDs for each object derived from a first fit. To these best-fit SPEDs, we add the observed noise drawn assuming a Gaussian distribution (see \citealt{Boquien2019} for a more detailed explanation).


\begin{figure}
    \begin{centering}
	\includegraphics[width=0.95\columnwidth]{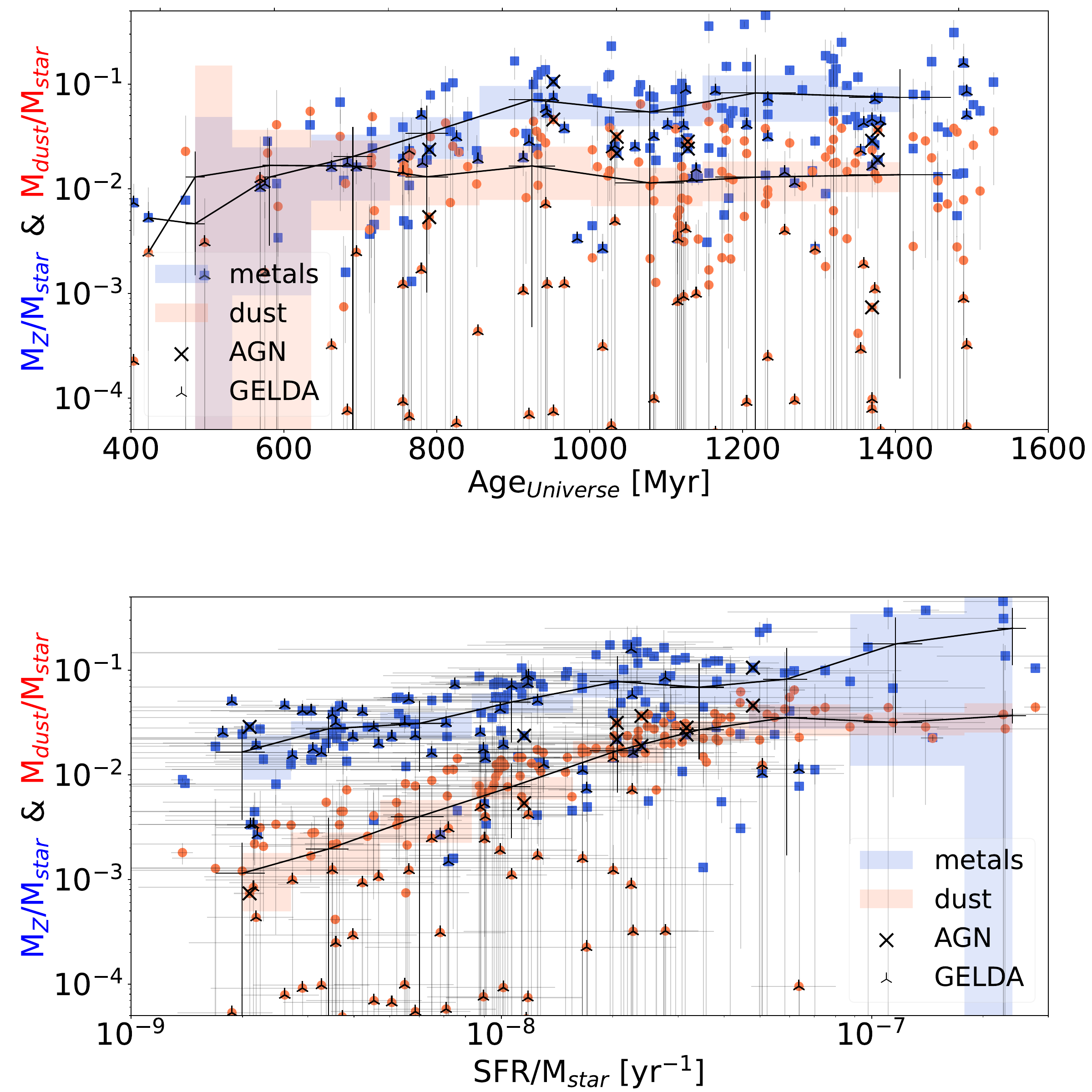}
    \caption{The evolution of the mass of metals and mass of dust: M$_Z$/M$_{star}$ (blue dots) and M$_{dust}$/M$_{star}$ (red dots) both regularly increase with the age of the Universe (top): more metals and dust grains are formed as the Universe ages. We note that even for galaxies (blue symbols) in the early Universe (age$_{Universe}$ $\lesssim$ 600 Myr or z$\gtrsim$9), M$_Z$/M$_{star}$ never goes below a few 10$^{-3}$, possibly suggesting a fast rise of metals that produces the observed threshold. M$_Z$/M$_{star}$ increases faster than M$_{dust}$/M$_{star}$, with a much larger dispersion for M$_{dust}$/M$_{star}$. The light blue/red area show the mean and 2$\sigma$ confidence interval of the distribution within several bins. b) M$_Z$/M$_{star}$ (in blue) and M$_{dust}$/M$_{star}$ (in red) decrease from high to low sSFR (bottom) as also shown in e.g. \protect\citealt{Palla2024,Shivaei2022}. We also note here a larger dispersion at lower sSFR for M$_{dust}$/M$_{star}$, but only for M$_{dust}$/M$_{star}$.}
    \label{fig:dust_metals}
    \end{centering}
\end{figure}

In Fig.~\ref{fig:dust_metals} we compare the trends related to the increase in the specific mass of metals (M$_Z$/M$_{star}$) and of dust (M$_{dust}$/M$_{star}$) with cosmic ages and with the star formation rate sSFR=SFR/Mstar. Several points can be noticed: first, both follow a similar trend; second, M$_Z$/M$_{star}$ is always higher than M$_{dust}$/M$_{star}$: the mass of metals is larger than the dust mass; Third, we observe a lack of extremely low-metallicity galaxies: all the galaxies observed so far are above a critical metallicity value (Bayesian values derived by CIGALE) of about Z$_{crit}$=10$^{-3}$.

A quantization of the mean dust-to-metal, DTM=<M$_{dust}$/M$_Z$> shown in Fig.~\ref{fig:dust_metals}, gives DTM$_{mean}$ = 0.080, DTM$_{median}$ = 0.006 for GELDAs with a first quartile (25 \%) Q1 = 0.002 and a third quartile (75 \%) Q3 = 0.037, whereas for non GELDAs DTM$_{mean}$ = 0.654, DTM$_{median}$ 0.156 with Q1 = 0.085 and Q3 = 0.331. We observe a strong break in the DTM that could be interpreted (\citealt{Inoue2011, Asano2013, Zhukovska2014, Feldmann2015, Popping2017, Hou2019, Li2019, Graziani2020, Triani2020, Parente2022, Choban2024, Dubois2024}) as hints that GELDAs have not started accretion growth of dust while non GELDAs are above the critical metallicity and have dust growth in the ISM. It is interesting to note that \cite{RemyRuyer2014} observed a large scatter in the gas-to-dust mass ratio for a sample of 126 galaxies spanning a 2 dex range in metallicity. This scatter appears at 7.2 $\lesssim 12+\log_{OH} \lesssim$ 8.7 and is consistent with the dust growth in the ISM predicted by \cite{Asano2013} and other works cited in this paper. The objects are those appearing on the bottom left of Fig.~\ref{fig:AFUV-Mstar} and Fig.~\ref{fig:dust_metals}, that is GELDAs. 

\section{Discussion}
\label{Discussion}

\begin{figure}
	\includegraphics[width=\columnwidth]{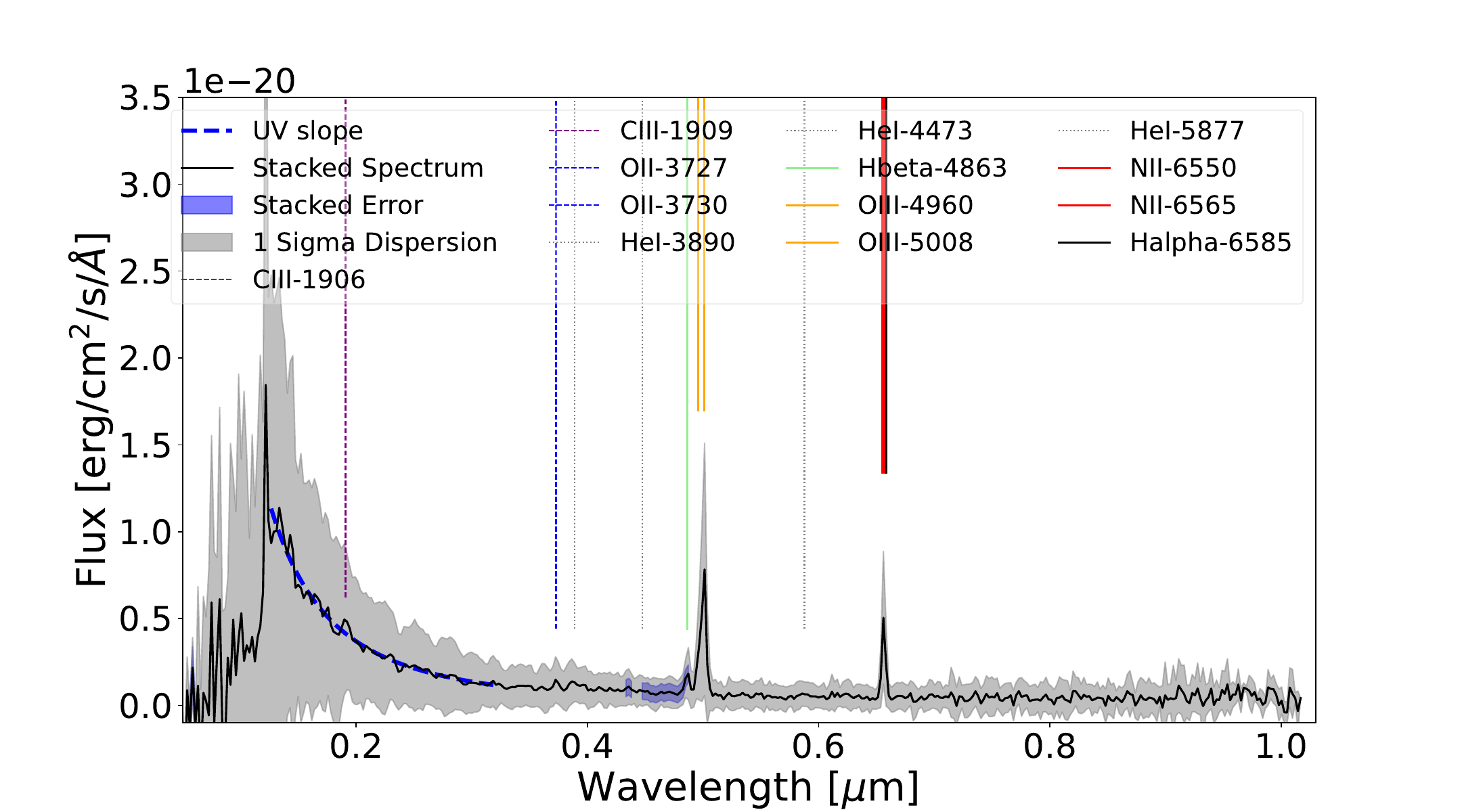}
    \caption{Stacked spectrum of the objects selected in the lower sequence. Vertical lines show the location of a few usual emission lines (see inside caption). We show the position of the lines, including where the HeI lines would be expected. None of them are detected, confirming these objects do not contain a dominant AGN. This stacked spectrum is similar to that of a young starburst, with a blue UV slope and prominent Hydrogen and [OIII) emission lines.}
    \label{fig:stack}
\end{figure}

We now analyze the possible origins of GELDAs. To this aim, we first stack the spectra of the GELDAs. See Fig.~\ref{fig:stack} and line fluxes extracted from the stacked spectrum in Tab.~\ref{Tab.stack}. This spectrum is characteristic of star-forming galaxies, with a very blue UV slope $\beta_{FUV}$ = -2.451 $\pm$ 0.066. This sample of GELDAs is compatible with no dust attenuation for Case B\footnote{from \cite{Groves2012}: "Case A and Case B. Case A assumes that an ionized nebula is optically thin to all Lyman emission lines, while Case B assumes that a nebula is optically thick to all Lyman lines greater than Ly$\alpha$, meaning these photons are absorbed and re-emitted as a combination of Ly$\alpha$ and higher order lines, such as the Balmer lines. These two cases will lead to different intrinsic ratios for the Balmer lines, with variations of the same order as temperature effects. Although Case B is typically assumed for determining intrinsic ratios, in reality the ratio in typical H~II regions lies between these two cases."}: H$\alpha$/H$\beta$=2.932±0.660. To reach H$\alpha$/H$\beta$=2.86 (no dust attenuation), we need to apply a correction to H$\beta$ for the underlying absorption of 2.5 \%, a relatively low correction (\citealt{Kashino2013, Reddy2015, Shivaei2020}).
 
\begin{table}
\begin{centering}
\caption{Measured UV slope and fluxes of the emission lines of the GELDA stacked spectrum are in erg/cm$^2$/s/$\AA$. We correct the H$\beta$ flux for the underlying absorption that increases H$\beta$ by 2.5 \% but we do not correct for H$\alpha$ which is assumed negligible (\protect\citealt{Kashino2013, Reddy2015, Shivaei2020}. With this underlying absorption we would conclude there is no dust attenuation for Case B for these GELDAs.}
\begin{tabular}{|l|r|r|}
\hline
  \multicolumn{1}{|c|}{Parameter} &
  \multicolumn{1}{c|}{Value} \\
\hline
flux CIII $\lambda$1907 $\AA$ & $<$ 7.956e-20 \\
flux CIII $\lambda$1909 $\AA$ & $<$ 7.956e-20 \\
flux OII $\lambda$3727 $\AA$ &  $<$  6.105e-21 \\
flux OII $\lambda$3730 $\AA$ &  $<$  6.105e-21 \\
flux OIII $\lambda$4960 $\AA$ & 7.977E-20 $\pm$  1.039e-20 \\
flux OIII $\lambda$5008 $\AA$ & 2.559E-19 $\pm$  1.039e-20 \\
flux H$\beta$ (measured) &  4.850E-20 $\pm$  1.039e-20 \\
flux H$\beta$ (2.5 \% abs.) &  4.971e-20 $\pm$  1.039e-20 \\
flux H$\alpha$&  1.422E-19 $\pm$  9.771e-21 \\
flux NII $\lambda$6550 $\AA$ &  $<$  9.771e-21 \\
flux NII $\lambda$6585 $\AA$ &  $<$  9.771e-21 \\
\hline
$\beta_{FUV}$& -2.451 $\pm$ 0.066 \\
$R_{H}$(F200W) & 380 $\pm$ 132 pc \\
$M_{FUV}$ & -18.9 $\pm$ 0.9 \\
$\log_{10}(M_{star}$) & 8.36 $\pm$ 0.37 \\
\hline
\end{tabular}
\label{Tab.stack}
\end{centering}
\end{table} 

The simplest origin is that these GELDAs are almost unaffected by dust attenuation because they did not produce a significant dust mass. Although this might be possible at z $\gtrsim$ 8.8, this is less likely at lower redshifts because of the global increase in the dust mass density that could pollute gas in the intergalactic medium (e.g. \citealt{Madau2014A, Traina2024} or in the average dust attenuation of galaxies (\citealt{Burgarella2013, Bogdanoska2020} from the early Universe to z=3-4. Moreover, some works (e.g. \citealt{Langeroodi2024}) suggest that dust is formed very fast by SNae on timescales shorter than $\sim 30$ Myr. Even if not all dust grains had been destroyed by the SNae reverse shock, some residual attenuation 0.05 $\leq$ A$_V$ $\leq$ 0.2, which translates to A$_{FUV}$ = 0.15 - 1.0 depending on the dust attenuation law (\citealt{Salim2020}), and should still be detectable.

\begin{figure}
	\includegraphics[width=0.9\columnwidth]{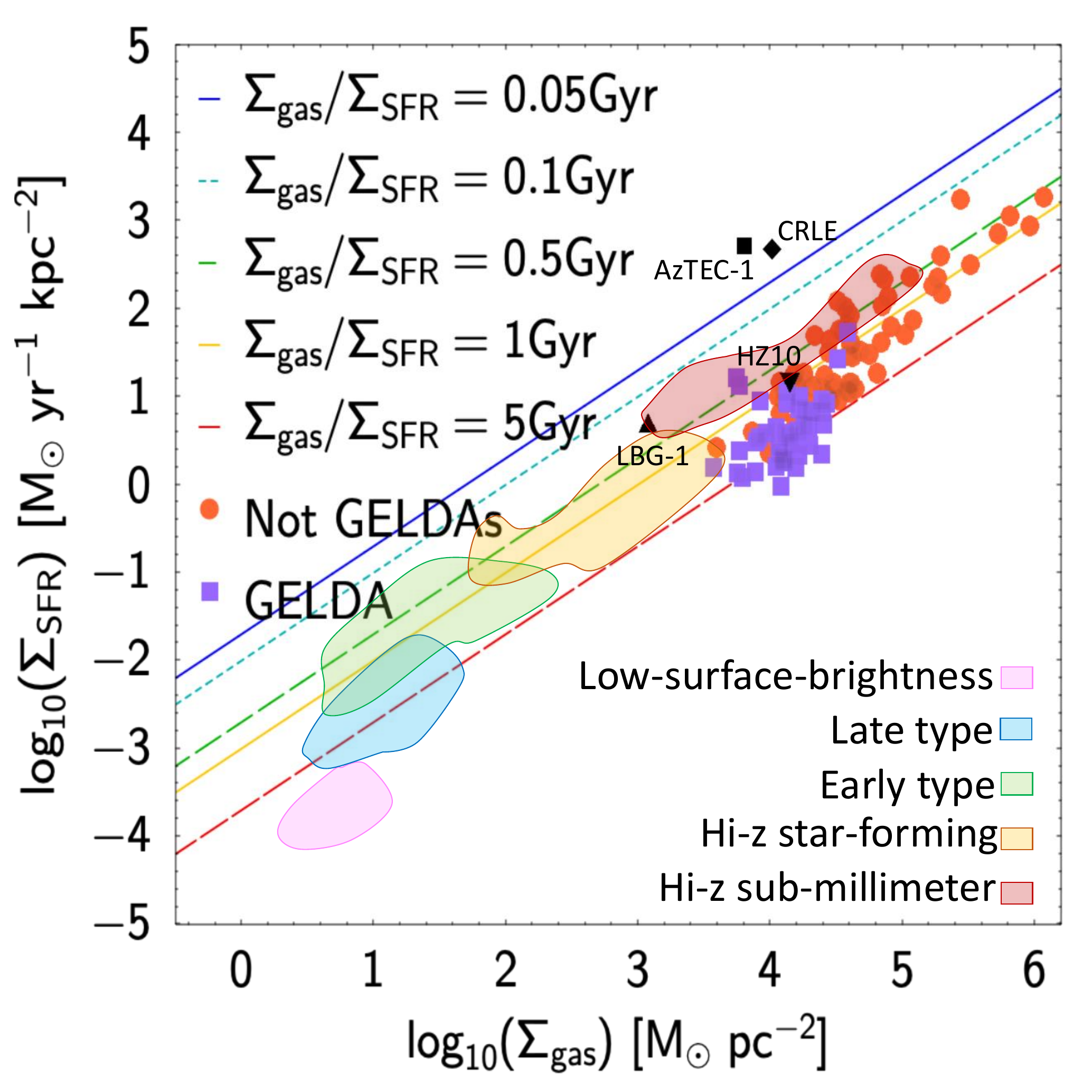}
    \caption{SFR surface density as a function of the gas mass surface density for a sample of local and high-redshift galaxies, including HZ10, LBG-1, AzTEC-3, and CRLE adapted from \protect\cite{Pavesi2019} and \protect\cite{Shi2011}. Our objects, GELDAs and not GELDAs are very dense, and all of them are found a bit below other objects, which might be related to the very high gas density in these objects where the turbulence can produce a negative effect on the SFE, which is thus lower than in other high-redshift galaxies at the same gas surface density. From bottom to top, the shaded areas correspond to low-surface-brightess (magenta), late type (blue), early type (green), hi-z star-forming (yellow) and hi-z sub-mm (red) galaxies. They are also given inside the figure.}
    \label{Fig_SigmaSFR_Sigmagas}
\end{figure}

Another origin could be related to the relative geometry of dust and stars in these objects or their small sizes. We know that the brightest H II regions in local galaxies show a correlation between the Balmer line reddening and the dust mass surface density (\citealt{Kreckel2013, Trayford2020, Seille2022, Robertson2024}). Our high redshift galaxies are small (Tab.~\ref{Tab.stack}). The half-light radii measured in the F200W NIRCam images are $<$RH$_{F200W}$$>$ = 380 $\pm$ 132 pc and even less for z $>$ 8.8: $<$RH$_{F200W}$$>$ = 327 $\pm$ 87 pc) and thus very dense. We measure the surface densities of gas $0\lesssim\log_{10}(\Sigma_{gas}[M_\odot pc^{-2}])\lesssim6$ and the surface densities of SFR $4\lesssim\log_{10}(\Sigma_{SFR}[M_\odot yr^{-1} kpc^2]\lesssim3$. Star formation in galaxies is closely related to the local gas density and follows the so-called Schmidt law (\citealt{Schmidt1959}). In the dense cores of star formation regions studied in the Milky Way (e.g. \citealt{Shimajiri2017, Mattern2024}) and in local spiral galaxies (\citealt{Gao2004}), active high-mass star formation is intimately related to the very dense molecular gas, M$_{dense}$. However, when the density of the gas reaches an H$_2$ surface density $\Sigma_{H2} \gtrsim 100-200 ~M_\odot pc^{-2}$ (\citealt{Mattern2024}), a density much lower than the estimated values for our galaxies, turbulence can partially prevent star formation because of turbulence and reduce the SFR. This could provide us with an explanation for the location of our sample at low star formation efficiency (SFE), below other objects at the same densities in Fig.~\ref{Fig_SigmaSFR_Sigmagas}, offset from both hi-z sub-mm galaxies. These low SFEs are not in agreement with the suggested high FFB-related SFE which is one of the suggested origins of the excess of UV-bright galaxies in the early Universe.

A third possible explanation can be found in \cite{Ferrara2023} where they built a model that reproduces the excess of UV-bright galaxies in the luminosity functions at z = 10-14. They propose that these galaxies at z$\gtrsim$11 would contain negligible amounts of dust and that most of the dust produced by SNae in these objects could have been efficiently ejected during the very first phases of galaxy build-up because these galaxies are bursty and they could temporarily evacuate large amounts of gas and dust far from the star-forming region (e.g. \citealt{Sun2023, Choban2024}). In this case, the lower objects could correspond to galaxies where dust is efficiently ejected far from the stellar populations by radiation pressure as soon as it is produced by stars. However, if winds had ejected dust in high-redshift galaxies, they probably also expelled gas. However, both for GELDAs and non-GELDAs, we find $f_{gas} = M_{gas}/(M_{gas}+M_{star})$ = 0.96 $\pm$ 0.03 in agreement with models predicting that galaxies with $\log_{10}(M_{star}/M_\odot)<9.0$ have f$_{gas}\gtrsim 0.75$ (\citealt{Dave2017, Popping2014}). However, galaxies at z $>$ 8.8 have slightly lower but still very high $f_{gas} = 0.90 \pm 0.05$. Thus, these gas fractions show that these objects still contain a large mass of gas and should therefore also contain dust.

\begin{figure}
    \begin{centering}
    \includegraphics[width=\columnwidth]{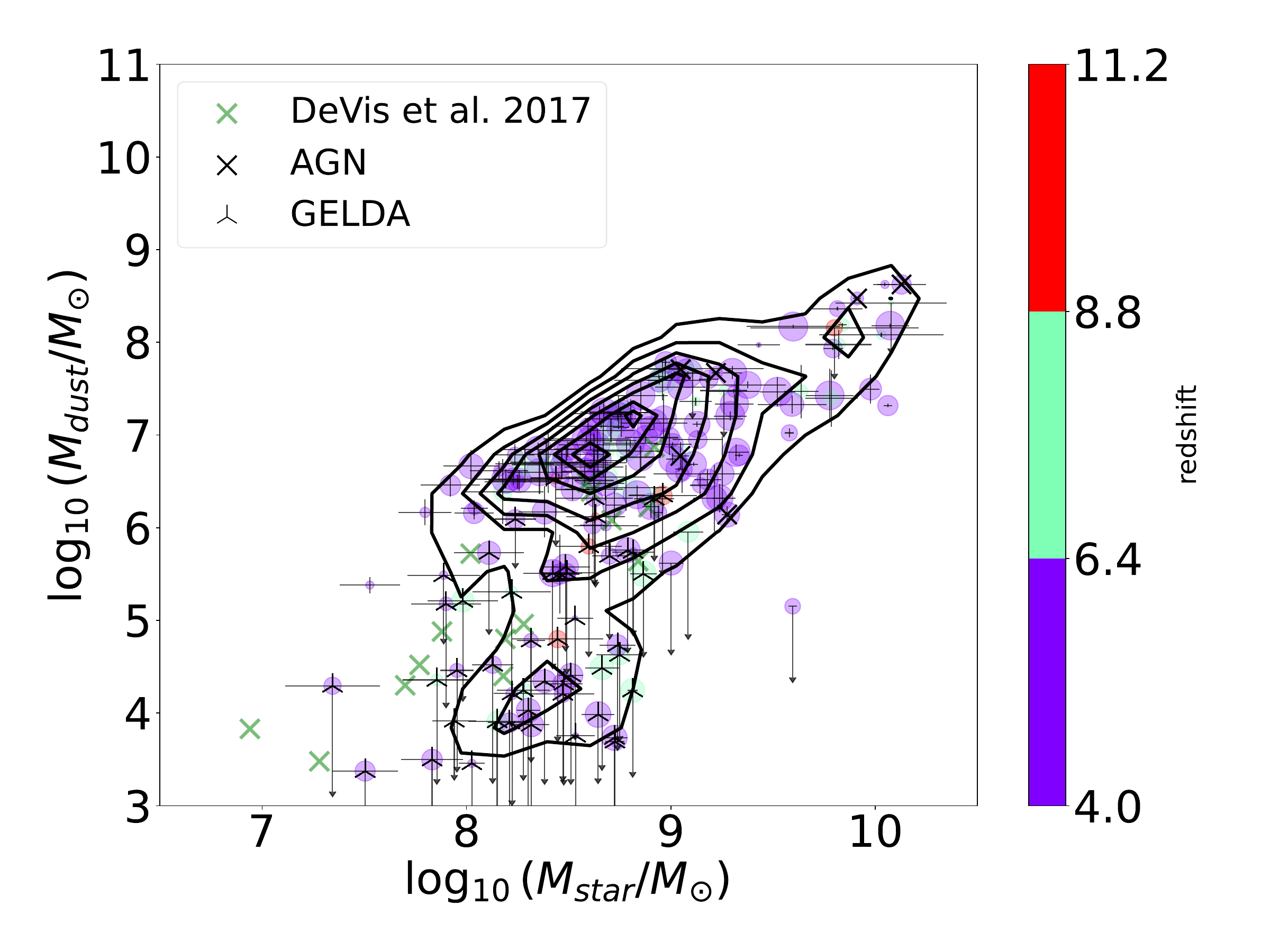}
    \includegraphics[width=0.8\columnwidth]{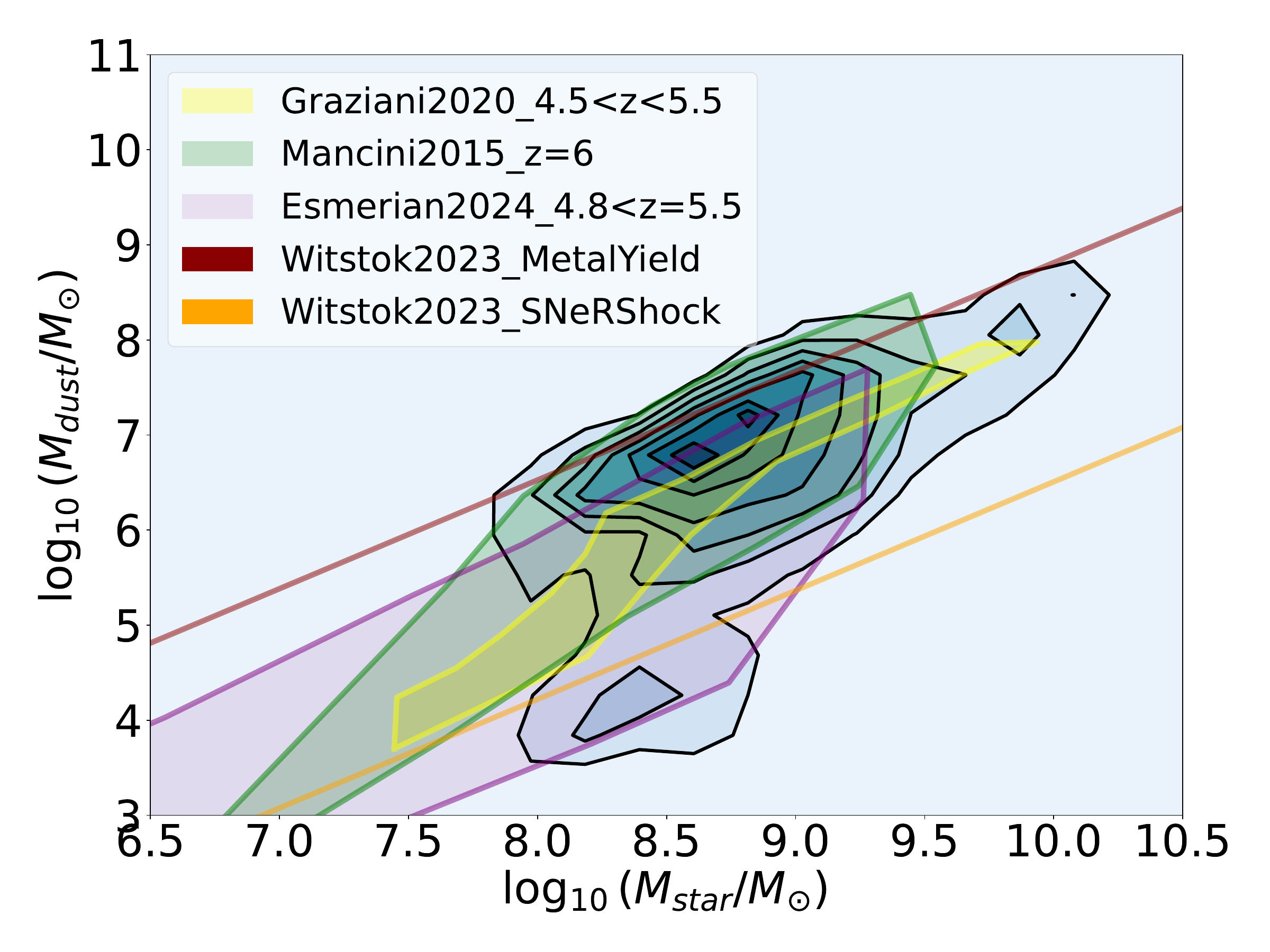}
    \caption{M$_{dust}$ as a function of M$_{star}$. Top: The dots are color-coded in redshift. The size of the symbols provides us with information on 12+$\log_{10}$(O/H) with the largest sizes corresponding to the largest metallicities. We show density contours in black, heavy lines. At $\log_{10}$ (M$_{star}$) $\sim$ 10$^{8-9}$ M$_\odot$, we observe a transition from an apparent high sequence to a lower one. The upper sequence is similar to that observed at low redshift (e.g. \protect\citealt{Beeston2018}). The lower objects have not been identified before in this (M$_{dust}$ vs. M$_{star}$) diagram, except for a few objects (green crosses, \protect\citealt{DeVis2017}). They share the same location in the plot and are extracted from a subsample of galaxies in an HI-selected sample from GAMA and H-ATLAS of local galaxies. Bottom: We plot models on the density contours: the hydrodynamical code dustygadget (\citealt{Graziani2020}) in yellow, a cosmological hydrodynamical simulation coupled with a chemical evolution model (\protect\citealt{Mancini2015}) in green and a suite of cosmological, fluid-dynamical simulations of galaxies (\protect\citealt{Esmerian2024}) in pink. The upper brown line (\protect\citealt{Witstok2023}) shows where galaxies with ISM-grown grains should be. The bottom orange line corresponds to galaxies with only stardust (\protect\citealt{Witstok2023}) that underwent a 95 \% destruction of grains by reverse SNae shock \citealt{Witstok2023}). The objects observed at the bottom of the top panel correspond to the transition between galaxies only containing stardust to galaxies dominated by ISM dust.}
    \label{fig:MdustMstar}
    \end{centering}
\end{figure}

Finally, in the M$_{dust}$ vs. M$_{star}$ diagram plotted in Fig.~\ref{fig:MdustMstar}, we observe a concentration of galaxies in the upper part of the figure, mainly in the range of 8.0$<$$\log_{10}$(M$_{star}$)$<$10.0. We also see a significant decline in dust attenuation at $\log_{10}$(M$_{star}$)$\sim$8.0-9.0 that was already seen in Fig.~\ref{fig:AFUV-Mstar}. This effect means that M$_{dust}$ is significantly lower by a factor of 100-1000 at a given stellar mass, with a lower clump or sequence, wel below the upper one. This biphasic plot could suggest a two-mode building of dust mass in galaxies. 

\begin{figure}
	\includegraphics[width=1.2\columnwidth]{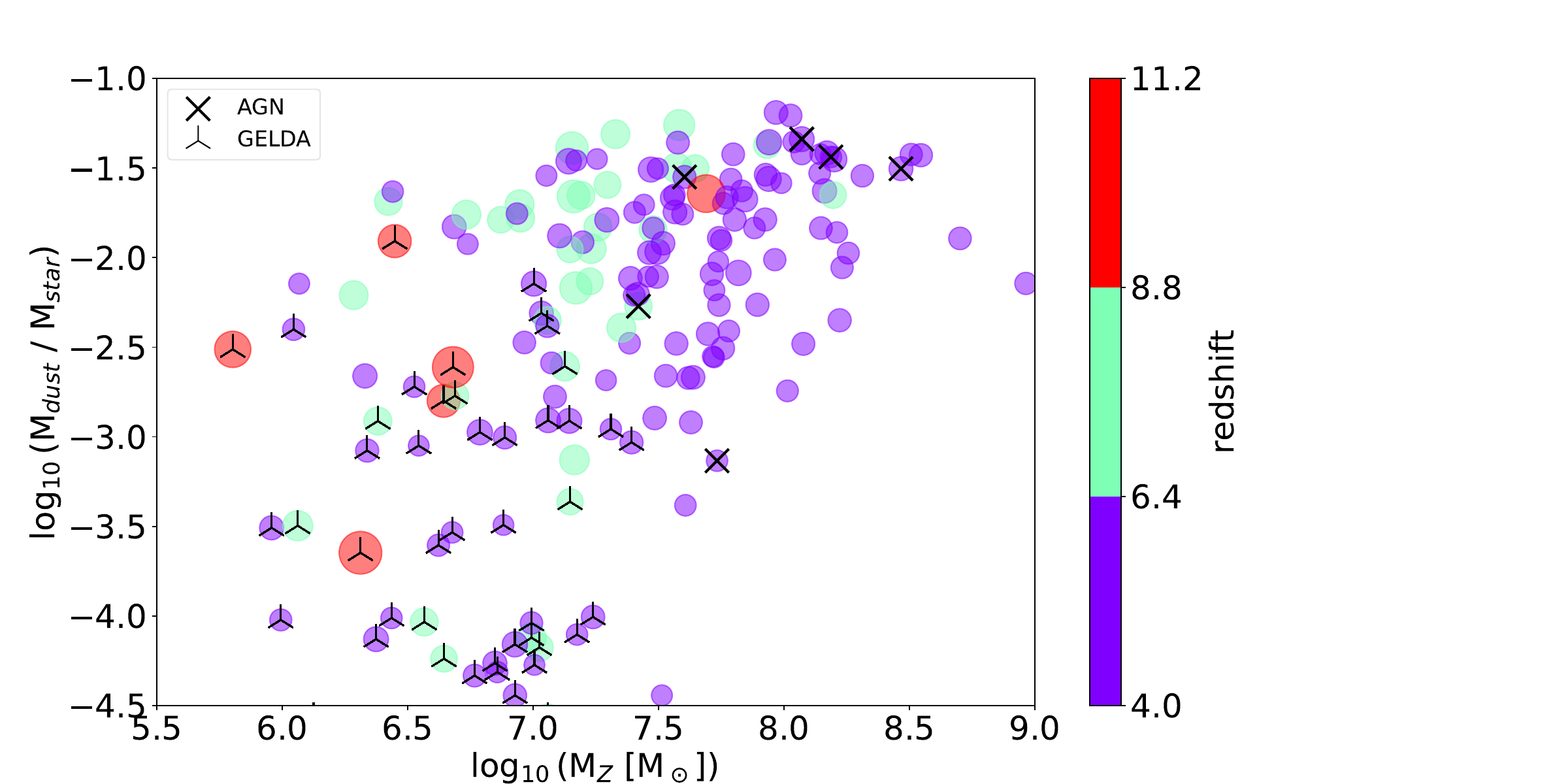}
    \caption{The objects are color-coded in redshift. AGN are shown as crosses and GELDAs are upside-down Y. This figure shows that GELDAs (both z $\geqslant$ 8.8 in red and z $<$ 8.8 in blue) have lower mass of metals than the rest of the sample. This would support the hypothesis that the origin of dust grains in these objects might not be due to ISM growth.}
    \label{fig:MdustMstar_MZ}
\end{figure}

To understand the nature of this lower sequence, Fig.~\ref{fig:MdustMstar} shows several models (\citealt{Mancini2015, Graziani2020, Esmerian2024, Witstok2023}). The first dust grains should have formed in stellar ejecta from SNae (and maybe AGB stars). However, a possibly substantial fraction of these dust grains is probably destroyed by the SNae reverse shock. After this first phase, the remaining dust grains form seeds and accrete ISM material for grain growth. This process seems to happen only when a critical ISM metallicity is reached at $0.05 \lesssim Z/Z_\odot \lesssim 0.5$ (\citealt{Inoue2011, Asano2013, Zhukovska2014, Feldmann2015, Popping2017, Hou2019, Li2019, Graziani2020, Triani2020, Parente2022, Choban2024}). While the upper sequence would have a dust mass where grains have grown in the ISM, the lower sequence would correspond to stardust grains only formed from SNae, with a grain destruction rate by the SNae reverse shock of the order of 95\% (\citealt{Witstok2023}). The jump from the lower to the upper sequence predicted by the models agrees well with our data. If this population of GELDAs only contains stardust, that we would provide a natural explanation for the excess of UV-bright galaxies at z$<$10 detected by JWST. We check in Figs.~\ref{fig:AFUV-Mstar-periodic} and \ref{fig:MdustMstar-others} that other assumptions lead to the same apparent transition. Only when no spectroscopic data is used in the fits does the shape of the transition change. Finally, Fig.~\ref{fig:MdustMstar_MZ} shows that there are significantly fewer metals in GELDAs compared to the rest of the sample. In this plot, the GELDAs are found in the bottom left of the figure with a metal mass $\log_{10}$(M$_Z$) $\lesssim$ 7.5 while the sample range extends 5.5 $\lesssim$ $\log_{10}$(M$_Z$) $\lesssim$ 9.0. This is expected if these objects did not undergo any growth of the dust grains triggered by a larger amount of metals in the ISM because this accretion of ISM material for grain growth is triggered when the metallicity reaches a minimum critical threshold (\citealt{Asano2013}).

\begin{figure*}
	\includegraphics[width=1.5\columnwidth]{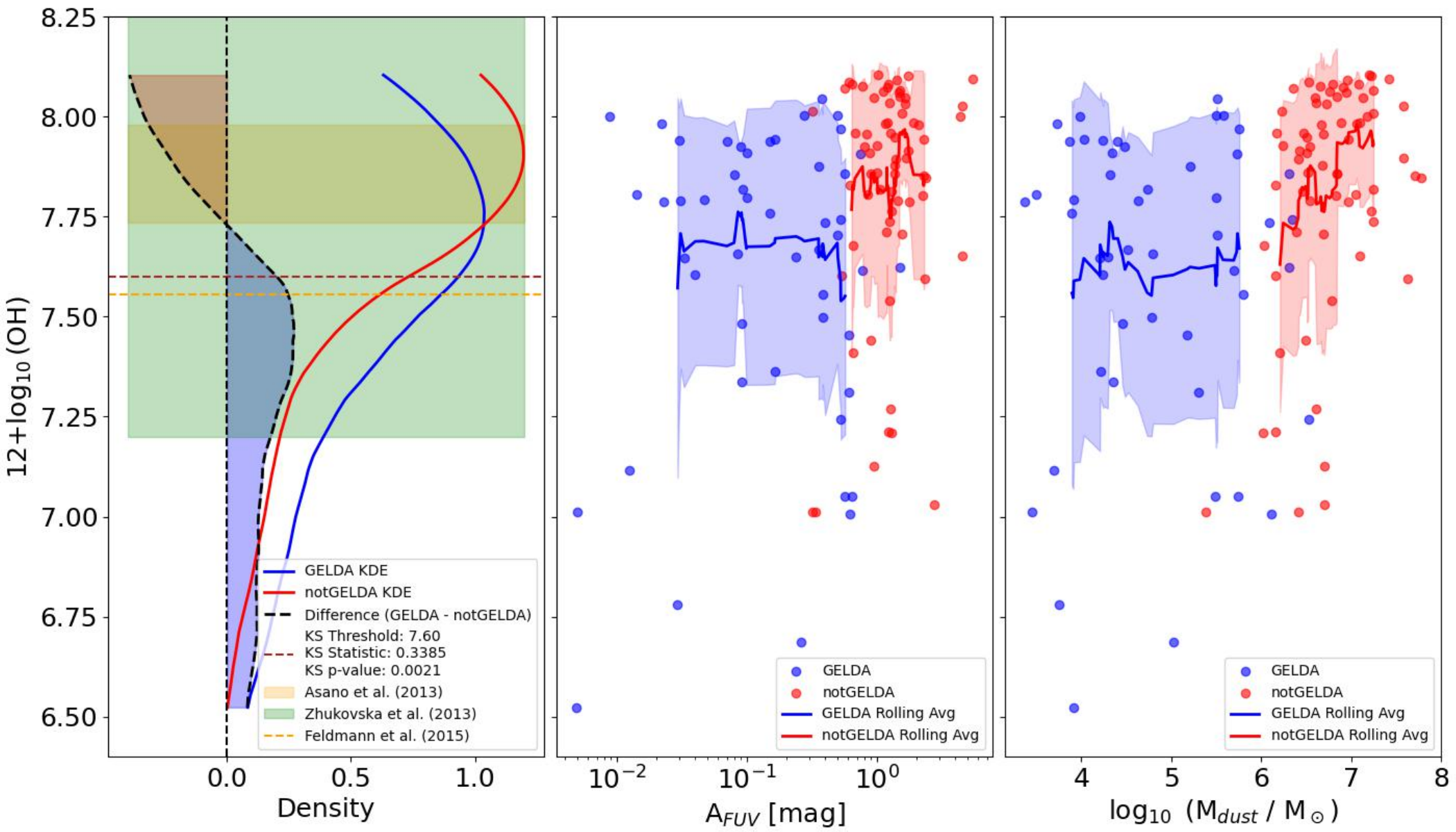}
    \begin{centering}
    \caption{Distribution of metallicity for GELDAs and not GELDAs. The left panel presents the kernel density estimation (KDE) with an epanechnikov kernel and a bandwidth=0.5 both for GELDAs (blue solid line) and non-GELDAS (red solid line). The black dashed line show the difference. The blue-shaded and red-shaded areas show where GELDAs and non-GELDAs are the dominant population, respectively. We clearly see that GELDAs preferentially cluster at low metallicity, with a threshold estimated at 12+$\log_{10}$(O/H)=7.60. The Kolmogorov-Smirnov confirms that the difference is highly significative. We show predictions for the transition from stardust to ISM from some of the models (see main text for more works): \cite{Asano2013} (orange-shaded area), \cite{Zhukovska2014} (green-shaded area) and \cite{Feldmann2015} (orange dashed line). \cite{RemyRuyer2014} found a larger scatter in their dust-to-gas ratio covers at almost the same metallicity range: 7.2 $\lesssim$ 12+$\log_{10}$(O/H) $\lesssim$ 8.5. All of these metallicities are in good agreement with our estimated threshold. The middle and right panels show the rolling averages (window size=15) of the metallicity as a function of A$_{FUV}$ and $\log_{10}$ (M$_{dust}$). Both panels confirm the difference between GELDAs (blue) and non-GELDAS (red).}
    \label{Fig1.12logOH}
    \end{centering}
\end{figure*}

To test whether our hypothesis could be correct, we try to check if we observe a difference in metallicity for GELDAs and not GELDas in Figs.~\ref{Fig1.12logOH}. The Kolmogorov-Smirnov (KS) test shown in Fig.~\ref{Fig1.12logOH} suggests that the difference between the two distributions (GELDAs and non-GELDAS) is highly significant. Furthermore, a metallicity threshold is found at 12+$\log_{10}$(O/H)=7.60, which corresponds (with Z$_\odot$=0.014 and Eq.~1) to Z=0.11. This is in excellent agreement with the critical metallicity (that is, the metallicity at which the contribution of stars equals that of the dust mass growth in the ISM) predicted by the models shown in Fig.~\ref{Fig1.12logOH}, in agreement with most models listed in this paper. For example \cite{Asano2013} give: Z/Z$_\odot$=0.2, that is 12+$\log_{10}$(O/H) = 7.86, and \cite{Feldmann2025} give Z/Z$_\odot$=0.1, that is 12+$\log_{10}$(O/H) = 7.56.

The present results provide us with an inherent explanation for the UV-bright tension in the early Universe if these galaxies contain only a low dust mass mainly formed in the circumstellar medium around SNae in the very first phases of star formation and not by accretion in the ISM. This hypothesis is further supported by the fact that in our total sample 5/6 galaxies, that is $\sim$ 83.3 \% of the z $\geqslant$ 8.8 galaxies are GELDAs whereas at z $<$ 8.8 only 38/167, that is, 22.8 \% are GELDAs, suggesting this type of galaxies could become dominant in the early Universe.

\section{Conclusions}

We detected a population of galaxies with extremely low dust attenuation (GELDAs) in the redshift range 4.0 $<$ z $<$ 11.4 using a new version of the CIGALE code that accepts both photometric and spectroscopic data. GELDAs are defined as follows: 
\begin{itemize}
    \item A$_{FUV}$ = 0 within 2$\sigma_{A\_FUV}$, that is no dust attenuation,
    \item M$_{star}$ $<$ 10$^9$M$_\odot$
\end{itemize}

The present galaxy sample shares most of its properties with the ALPINE one in terms of stellar mass: 8.0$\lesssim$M$_{star}$$\lesssim$11.4 and redshift: two redshift ranges at 4.40$<$z$<$4.65 and 5.05$<$z$<$5.90 (\citealt{Burgarella2022}). Assuming that the far-IR dust emission of these GELDAs is similar to that of the ALPINE galaxy sample, we estimate dust masses. In the M$_{dust}$ vs. M$_{star}$ diagram, we clearly see a transition at $\log_{10}$(M$_{star}$) $\sim$ 8.5 between an upper and a lower sequence. A comparison with models suggests that the transition galaxies could mark the shift from dust solely produced by stellar evolution (stardust galaxies) to dust growth in the ISM of galaxies. The dust-to-metal ratios are very low for GELDAs: DTM$_{mean}^{GELDA}=0.080$, DTM$_{median}^{GELDA}=0.006$ with a third quartile Q3 = 0.037 and quite high DTM$_{mean}^{non-GELDA}=0.654$, DTM$_{median}^{non-GELDA}=0.156$ with Q3 = 0.331 for non-GELDA, in agreement with the hypothesis of stardust vs. ISM dust.

In our data, a KS test suggests this transition to appear at 12+$\log_{10}$(O/H)=7.60 (Z/Z$_\odot$=0.1), which is in excellent agreement with the predicted metallicity at which the contribution of stars would equal that of the growth of the dust mass in the ISM in \cite{Asano2013}.

The fraction of gas mass $f_{gas} = M_{gas}/(M_{gas}+M_{star})$ $>$ 0.9 for our entire sample of galaxies including GELDAs at all redshifts. This suggests that there is a large gas mass in the galaxies that was not expelled, and supports the hypothesis that dust formed in the galaxies should still remain inside the galaxies.

Finally, the SFE of our galaxies is in agreement with the Schmidt-Kennicutt law (\citealt{Kennicutt}), although at lower SFEs than high-redshift sub-mm galaxies at the same density of gas mass.

In the highest redshift bin at z$>$8.8, almost all galaxies ($\sim$ 83.3 \%) can be assigned to the GELDA category, while less than 1/4 of the low-redshift (z$<$8.8) galaxies are GELDAs. These different regimes might mark a transition around z $\sim$ 9. In this highest redshift bin, galaxies with low M$_{dust}$/M$_{star}$ and blue UV slopes contain young, metal-poor stars that may be forming their first dust grains from Pop. II and at z$>$9, possibly Pop. III stars, along with their first metals. 

Such stardust galaxies would be ideal suspects to produce the excess of UV-bright galaxies in the early Universe because they might become dominant in the early Universe.

We do not detect any extremely low-metallicity values above z$_{gas} \sim 10^{-3}$ in our sample of galaxies, and even at z$\gtrsim$8.8, suggesting either a bias in our sample or a rapid rise of metals in the early Universe.

\begin{acknowledgements}
DB and VB thank the Programme National Cosmology and Galaxies (PNCG) and the Centre National d'Etudes Spatiales (CNES) for their financial support.
\end{acknowledgements}

\bibliographystyle{aa} 
\bibliography{MetalsDustCoev}

\begin{appendix} 
\section{Description of the spectro-photometric CIGALE}
\label{appendix:CIGALE}

The concept of CIGALE was developed in the original paper (\citealt{Burgarella2005}) where multi-wavelength data from the far-UV to the far-IR could be used to derive physical parameters by fitting photometric SEDs. The present open and public version 2025.0, 17 January 2025 of CIGALE (\citealt{Boquien2019}) is written in Python and parallelized. It also includes a much larger number of modules (that is, physical processes and models) that provide the user with a rich choice to adapt the modeling phase to most galaxies. This Python CIGALE code is also one of the fastest SED fitting codes in the world (\citealt{Burgarella2024}), making it faster than some of the machine-learning-based codes (namely the convolutional neural network and the deep learning neural network as used in \citealt{Euclid2023}). However, the most important difference of this new version is the possibility to combine spectroscopic data to photometric data (hereafter spectrophotometric data or SPED) with their own uncertainties, while conserving CIGALE’s ability to fit several thousands of objects in a reasonable time. This means that whatever the parameters derived via the fitting process are, these parameters have to be consistent with both photometric and spectroscopic data. Fitting the 173 galaxies using 800 million models from this sample takes about 12 hours on a 48-core computer with 512 GB of memory, utilizing about 60 GB for each run.

In order to combine the two above data types, we have to normalize the spectrum to the photometry. We provide three options: 1) no normalization: raw data are combined, 2) we integrate the modeled spectra into the filters and estimate a global normalization factor through a $\chi^2$ when the signal-to-noise ratio $>$5 for the photometric bands used to compute $\chi^2$, and 3) we determine a wavelength-dependent normalization. We stress that normalizing the spectroscopic data to the photometric data could be problematic if the emission inside the photometric aperture is physically different from the emission inside the spectral slit. In this case, the resulting fit might not be realistic because of the different natures of the emitting regions. For instance, a dusty galaxy might present a clear region in the outskirts that could dominate the spectrum in UV but not elsewhere in the spectrum if both observations are not at the same position. Converging would thus be difficult, and a good global fit would not be reached. We therefore recommend being careful when combining the spectroscopic and photometric data. For small galaxies like ours, this issue is minimized because we are more likely to observe the same region photometrically and spectroscopically.

In order to simultaneously fit all the data, we need to inform CIGALE about the (sometimes wavelength-dependent) spectral resolution of the spectrometer to create resolution elements corresponding to the instrumental spectral resolution where the models are integrated. This phase is transparent to the user and is performed during the configuration of the CIGALE environment. A typical CIGALE spectrophotometric run appears similar to a photometric run from the user's point of view. 

\begin{figure}[h]
    \includegraphics[width=\linewidth]{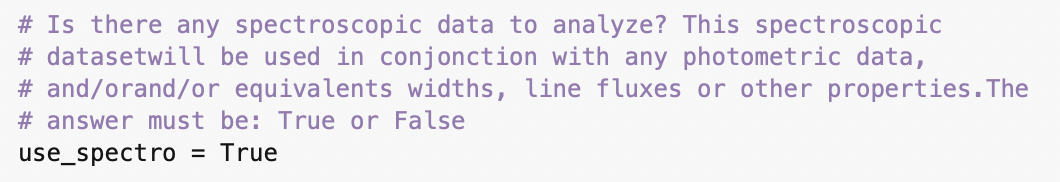}
    \caption{This flag must be set to "True" in the pcigale.ini file to fit spectroscopic data.}
    \label{fig:use_spectro}
\end{figure}

CIGALE learns that some spectroscopic data have to be taken into account from the configuration file, pcigale.ini, where a specific flag is set to 'True' as shown in Fig.~\ref{fig:use_spectro}.

Moreover, the input table must contain the following information:
\begin{itemize}
\item "id" that contains an alphanumeric identifier (a different one for each object to be fitted)
\item "redshift" that contains the redshift of the objects or "NaN" if redshifts have to be estimated.
\item "spectrum" which contains the path to the spectrum
\item "mode" that contains the type of spectrum, e.g. "prism" if JWST/NIRSpec data is used 
\item "norm" where one of the three normalizations should be provided for each object: "none", "global" or "wave". Note that the normalization could be different for each object. If no photometric data are given to CIGALE, "none" should be used, and only the spectrum will be fitted.
\end{itemize}

\section{Parameters used in CIGALE's final fit}
\label{sec:pcigale.ini}

\begin{figure}
\includegraphics[width=\columnwidth]{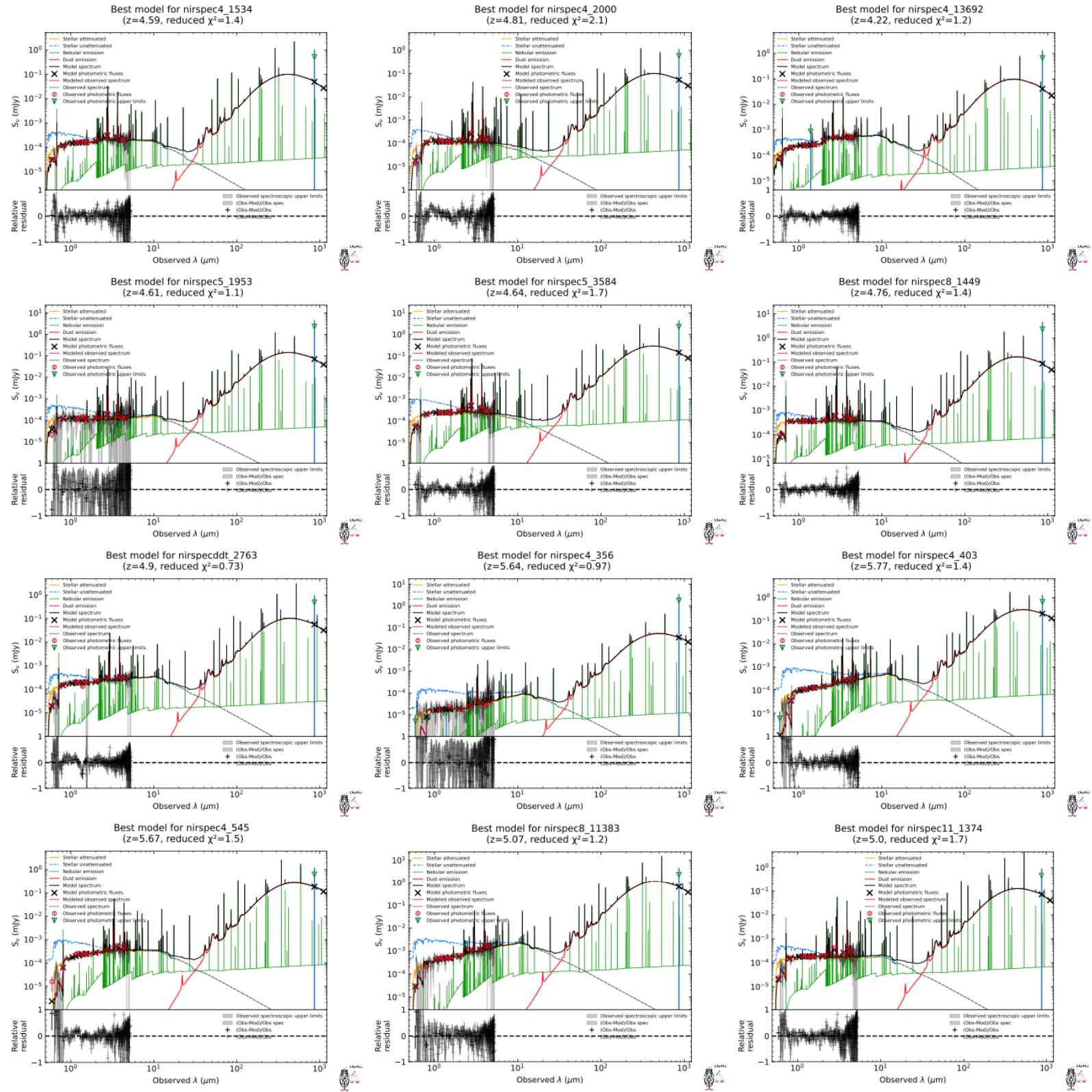}
    \caption{a) Objects in the upper sequence (larger M$_{dust}$) in M$_{dust}$ vs. M$_{star}$ plot. We present a sample of spectral fits over the whole spectral range, that is including NIRSpec spectroscopy and the sub-millimeter data (mostly upper limits). }
    \label{fig:SEDs_1a}
\end{figure}

\begin{figure}
\includegraphics[width=\columnwidth]{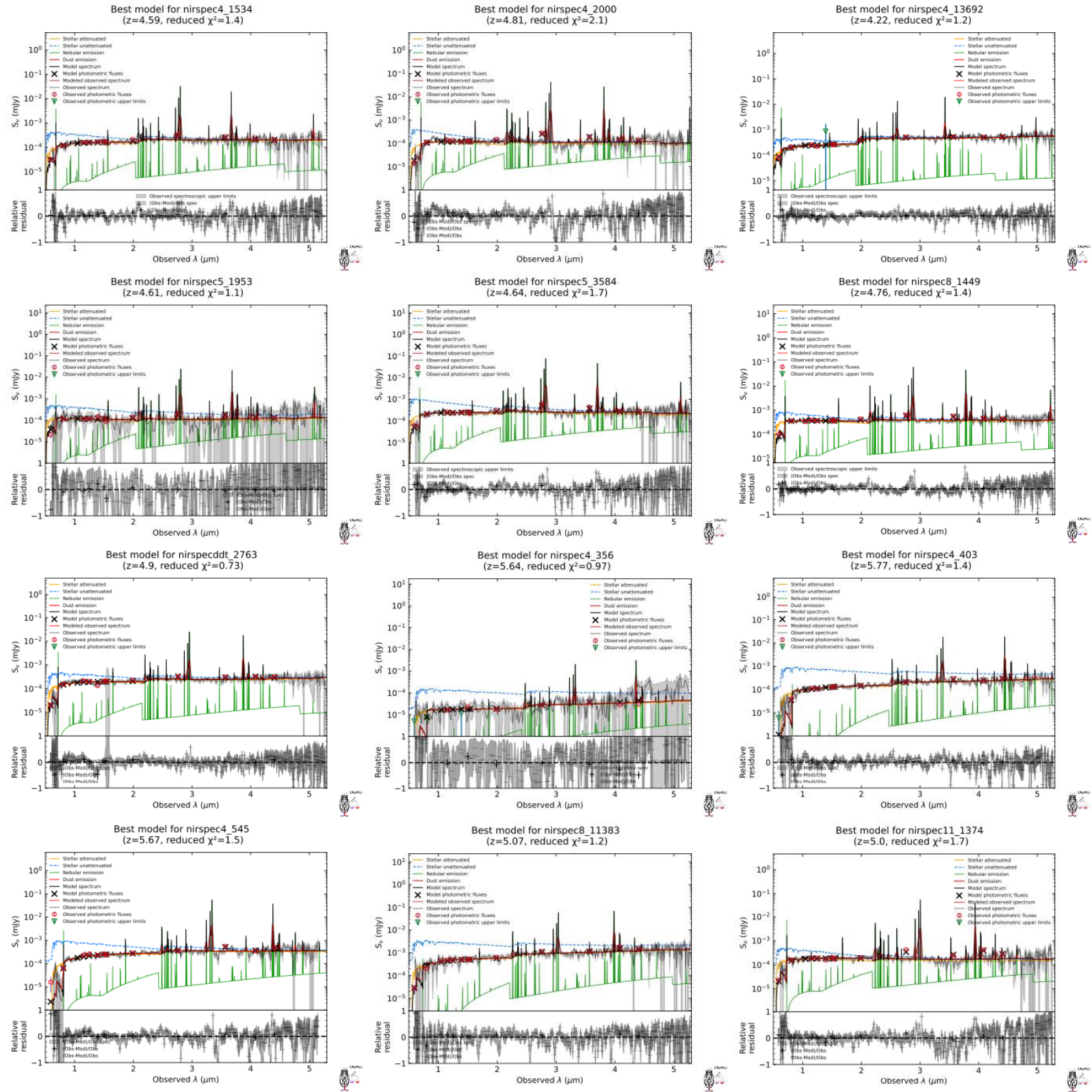}
    \caption{b) Same as Fig.~\ref{fig:SEDs_1a} for the same objects but only the fits of the NIRSpec spectrum is shown, which is a zoom in the previous plots.}
    \label{fig:SEDs_1b}
\end{figure}

\begin{figure}
\includegraphics[width=\columnwidth]{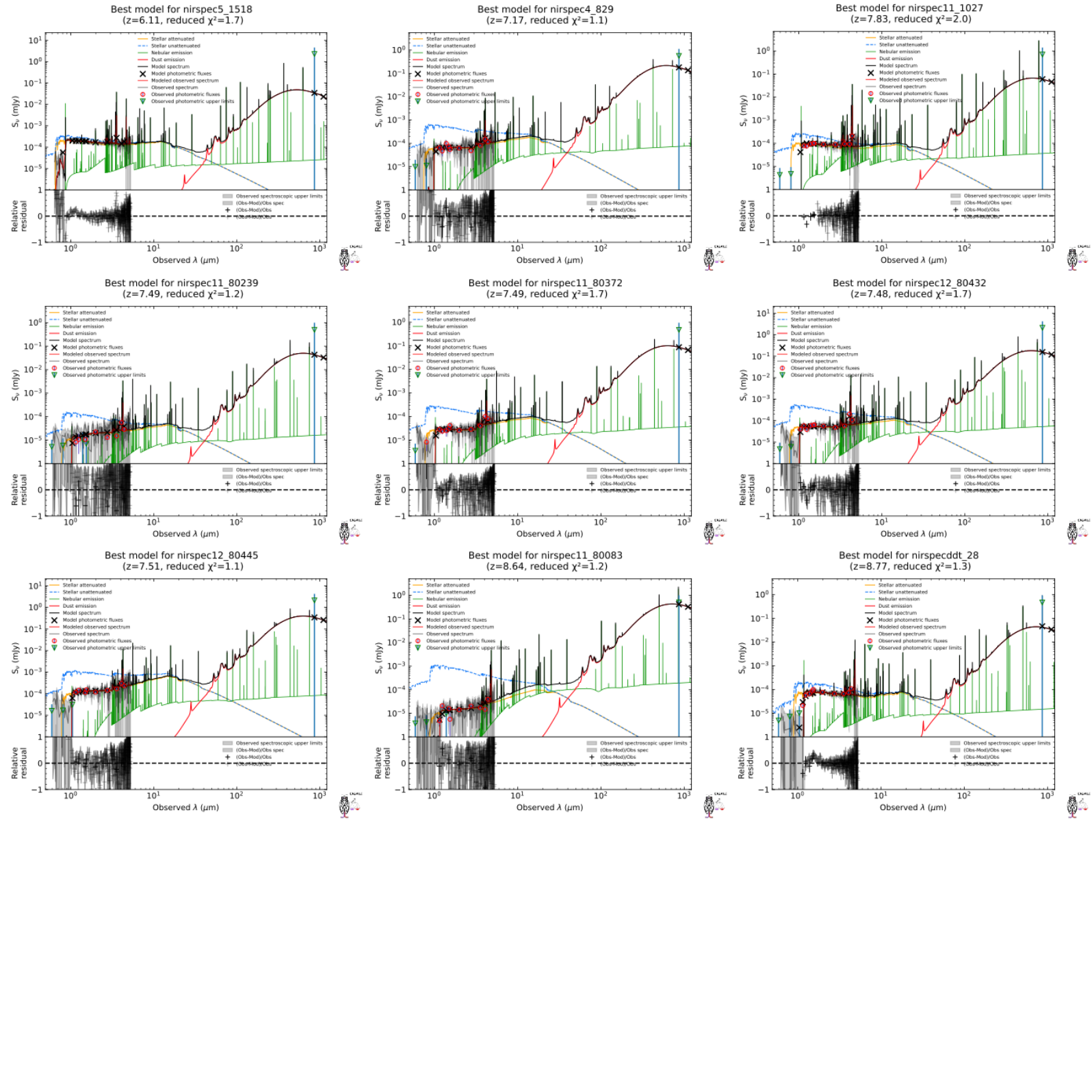}
    \caption{Same as previous Fig.~\ref{fig:SEDs_1a}.}
    \label{fig:SEDs_1c}
\end{figure}

\begin{figure}
\includegraphics[width=\columnwidth]{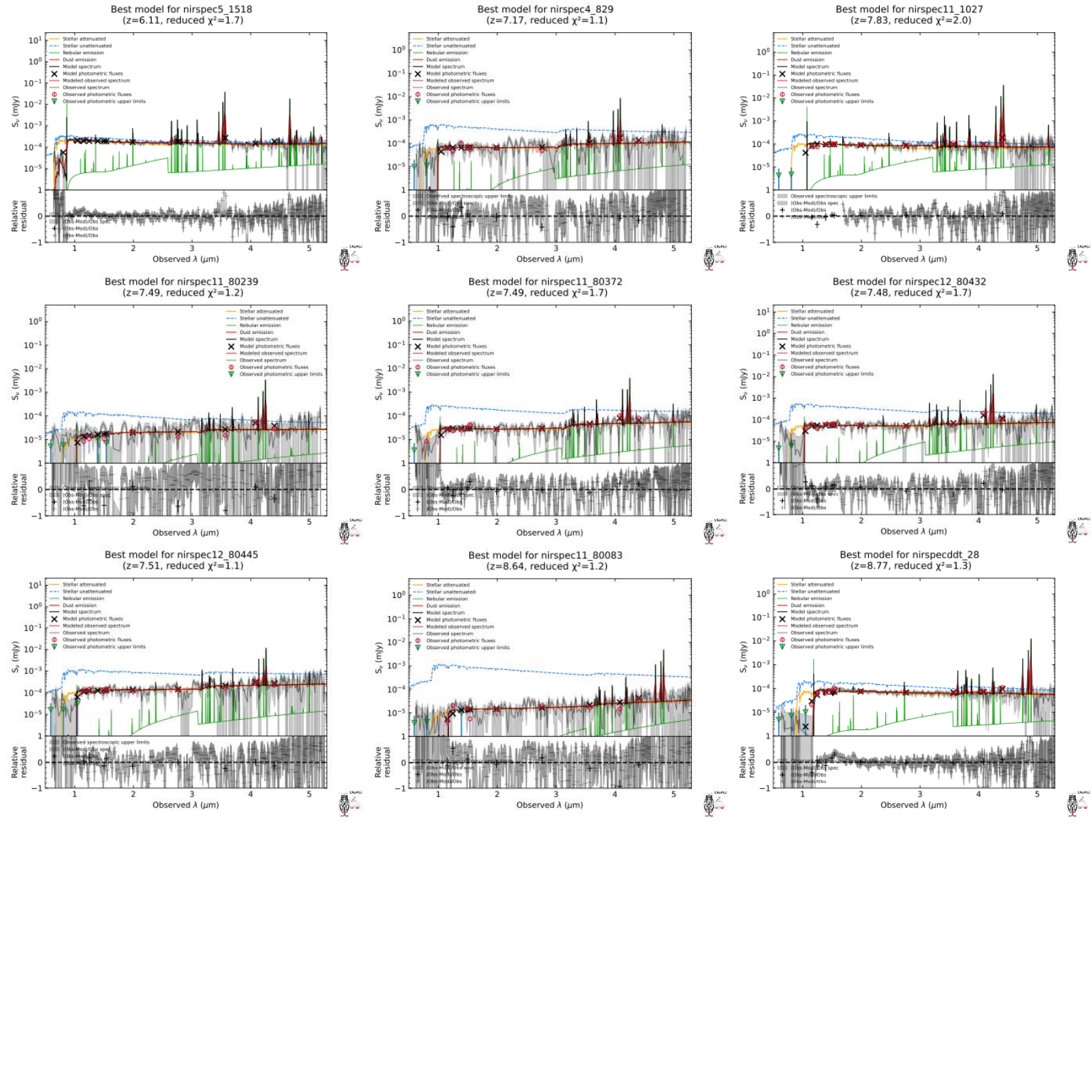}
    \caption{Same as Fig.~\ref{fig:SEDs_1b}.}
    \label{fig:SEDs_1d}
\end{figure}

\begin{figure}
\includegraphics[width=\columnwidth]{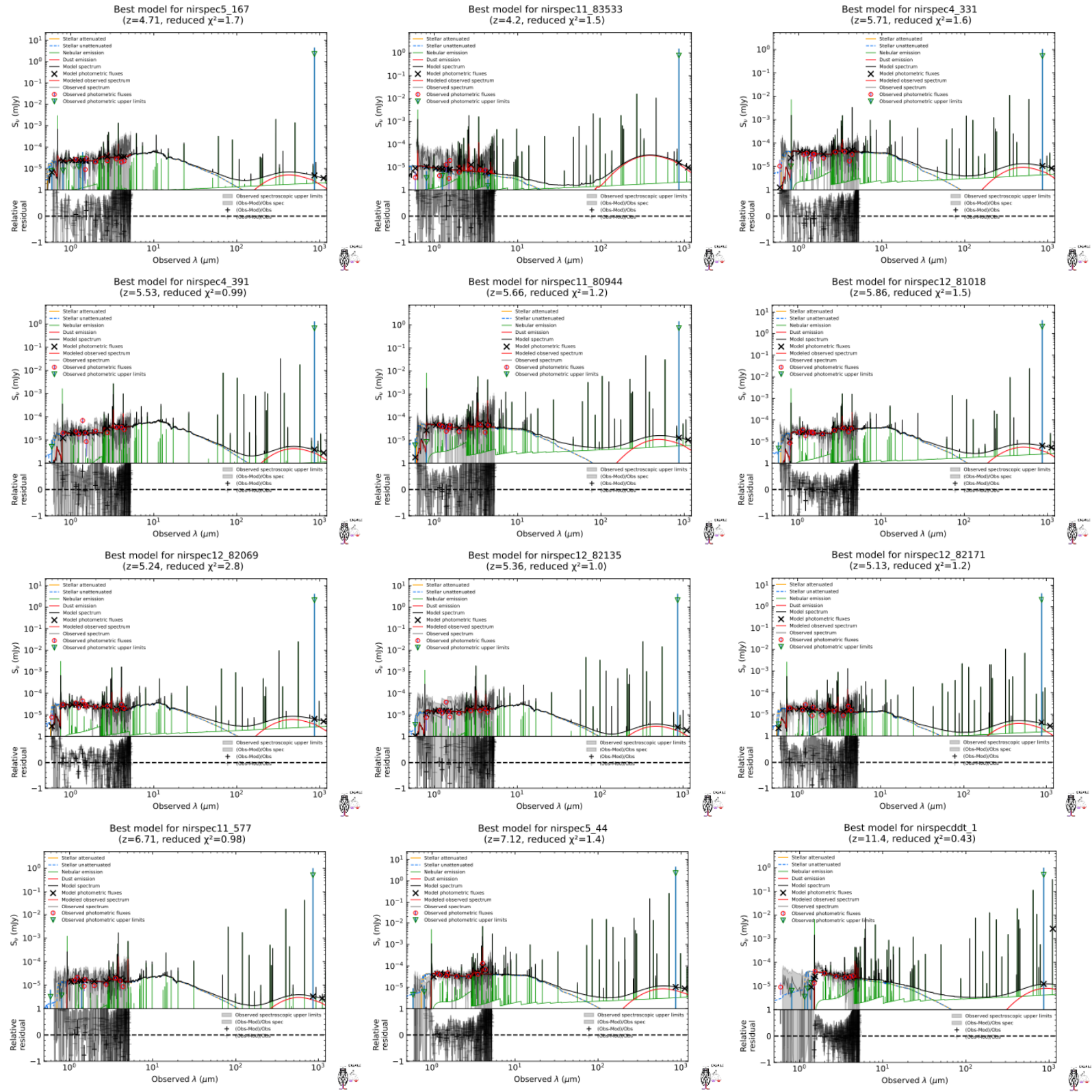}
    \caption{Objects in the lower sequence (lower M$_{dust}$) in M$_{dust}$ vs. M$_{star}$ plot. Same as Fig.~\ref{fig:SEDs_1a} but for objects in the lower sequence in the M$_{dust}$ vs. M$_{star}$ plot.}
    \label{fig:SEDs_2a}
\end{figure}

\begin{figure}
\includegraphics[width=\columnwidth]{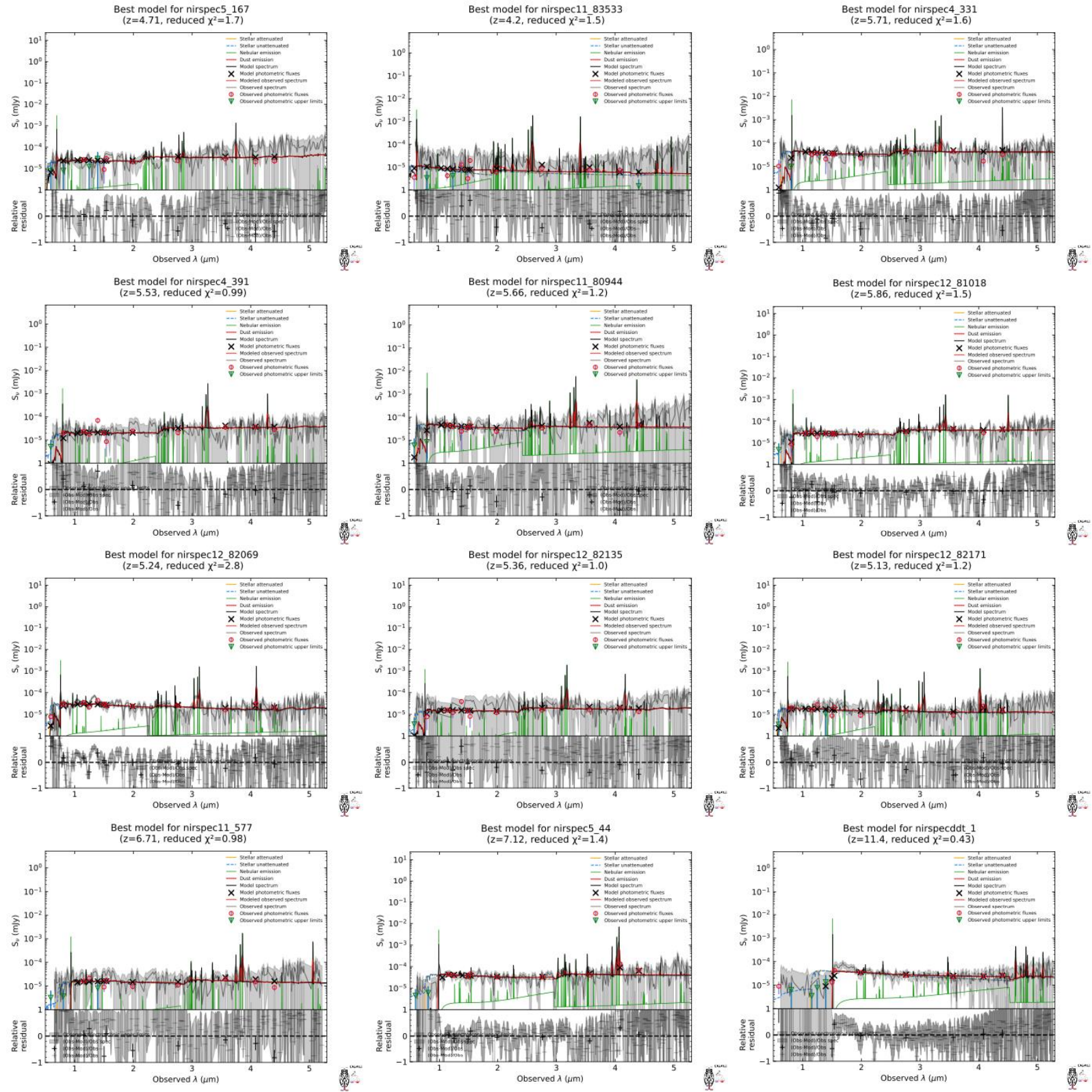}
    \caption{Same as Fig.~\ref{fig:SEDs_2a} for the same objects but only the fits of the NIRSpec spectrum is shown, which is a zoom in the previous plots.}
    \label{fig:SEDs_2b}
\end{figure}

All the spectral models computed by this new version of CIGALE (momentarily dubbed CIGALE-SPEC) using the selected modules (each one corresponding to a physical emission) are convolved with NIRSpec's prism spectral resolution matched to the observed spectra. The modules and the list of priors are listed in Tab.~\ref{Tab.pcigale.ini}. These spectral data are added to the photometry to form a SPED. The rest of the process follows the usual flow of CIGALE as described in \cite{Boquien2019}. The nebular models have been computed with CLOUDY, as described in \cite{Theule2024}. In this analysis, we normalize the prism spectroscopic data by computing a wavelength-dependent normalization, which is estimated from the photometry in the wavelength range, for the photometric bands that have a SNR$>$5.0. A one-dimensional piecewise linear interpolation with given discrete photometric data points is used to derive the wavelength-dependent normalization factor. This normalization is computed for each and every object with photometric data and applied to each spectroscopic observation. In this work, we used the WMAP7 cosmology (\citealt{Komatsu2011A}). We assumed a Chabrier initial mass function (IMF, \citealt{Chabrier2003}) with lower and upper mass cutoffs M$_{low}$ = 0.1 M$_\odot$ and M$_{up}$ = 100
M$_\odot$, and a solar metallicity Z$_\odot$ = 0.014 (\citealt{Bruzual2003}), and the dust emission models, use $\kappa_{\nu}$ = 0.637 m$^2$ kg$^{-2}$. We show examples of SPED fits in Figs.~\ref{fig:SEDs_1a}, \ref{fig:SEDs_1c} and \ref{fig:SEDs_2a} over the full spectral range and \ref{fig:SEDs_1b}, \ref{fig:SEDs_1d} and \ref{fig:SEDs_2b} over the NIRSpec range.

\begin{table*}[htp]
\begin{center}
 \resizebox{\linewidth}{0.5\height}{
\begin{tabular}{c|c|c|c} 
  \hline\hline
  {\bf Parameters} & {\bf Symbol} & {\bf Run \#1} & {\bf Run \#2} \\
  \hline
 Target  sample    &                      &  173 NIRSpec-observed objects & 173 NIRSpec-observed objects \\ 
  \hline\hline
    \multicolumn{4}{c}{}\\
    \multicolumn{4}{c}{\bf Delayed SFH and recent burst}\\
  \hline
 e-folding time scale of the delayed SFH & $\tau_{main}$ [Myr] & 500 & --- \\ 
  \hline
 Age of the main population & Age$_{main}$[Myr]  & 2, 5, 10, 25, 50, 250, 500, 1000 & ---  \\ 
  \hline
 e-folding time scale of the delayed SFH & $\tau_{burst}$ [Myr] & 10000 & --- \\ 
  \hline
 Age of the young population & Age$_{burst}$[Myr] & 1 & ---  \\ 
  \hline
    \multicolumn{4}{c}{}\\
    \multicolumn{4}{c}{\bf Periodic SFH}\\
  \hline
 Type of the individual star formation episodes & --- & --- & delayed \\ 
  \hline
 Elapsed time between the beginning of each burst & $\delta_{burst}$ [Myr] & --- & 10, 50, 100, 250 \\ 
  \hline
 e-folding time scale of the delayed SFH & $\tau_{burst}$ [Myr] & --- & 10, 100 \\ 
  \hline
 Age of the episodes & Age [Myr]  & --- & 1, 5, 10, 50, 100, 200, 400, 600, 800, 1000  \\ 
  \hline
    \multicolumn{4}{c}{}\\
    \multicolumn{4}{c}{\bf SSP}\\
  \hline
  SSP &   & BC03 & BC03 \\ 
  \hline
  Initial mass function &  IMF & Chabrier & Chabrier \\ 
  \hline
  Metallicity     & Z & 0.0001, 0.0004, 0.004, 0.008, 0.02, 0.05 & 0.0001, 0.0004, 0.004, 0.008, 0.02, 0.05 \\ \hline
    \multicolumn{4}{c}{}\\
    \multicolumn{4}{c}{\bf Nebular emission}\\
  \hline
  Ionization parameter &  logU    & -1.0, -1.2, -1.4, -1.6, -1.8, -2.0, -2.2, -2.4, -2.6, -2.8, -3.0, -3.2, -3.4, -3.6, -3.8, -4.0 & -1.0, -1.2, -1.4, -1.6, -1.8, -2.0, -2.2, -2.4, -2.6, -2.8, -3.0, -3.2, -3.4, -3.6, -3.8, -4.0 \\  \hline
  Gas metallicity &  Z$_{gas}$    & 0.0001, 0.0004, 0.001, 0.002, 0.0025, 0.003, 0.004, 0.005 & 0.0001, 0.0004, 0.001, 0.002, 0.0025, 0.003, 0.004, 0.005 \\  \hline
  Electron density &  n$_H$    & 10, 100, 1000 & 10, 100, 1000 \\
  Line width [km/s]    &     ---    &  150 &  150 \\
  \hline
    \multicolumn{4}{c}{}\\
    \multicolumn{4}{c}{\bf Dust attenuation law}\\
  \hline
  Color excess for both the old and young stellar populations &  E\_BV\_lines & 1e-4, 0.001, 0.010, 0.050, 0.10, 0.15, 0.20, 0.25, 0.30, 0.50, 0.75, 1.0, 1.5, 2.0, 3.0, 5.0 & 1e-4, 0.001, 0.010, 0.050, 0.10, 0.15, 0.20, 0.25, 0.30, 0.50, 0.75, 1.0, 1.5, 2.0, 3.0, 5.0 \\ 
  \hline
  Reduction factor to apply on E\_BV\_lines to compute E(B-V)s the stellar continuum attenuation. &  E\_BV\_factor &  0.44 & 0.44 \\ 
  \hline
  Bump amplitude &  uv\_bump\_amplitude &  0.0 & 0.0 \\ 
  \hline
  Power law slope & power law\_slope & -0.6, -0.30, 0.0, 0.30, 0.6 & -0.6, -0.30, 0.0, 0.30, 0.6 \\ 
  \hline
    \multicolumn{4}{c}{}\\
    \multicolumn{4}{c}{\bf Dust emission (DL2014)}\\
  \hline
  Mass fraction of PAH & $q_{PAH}$ & 0.47 & 0.47 \\ 
  \hline
  Minimum radiation field &  U$_{min}$ & 17.0 & 17.0 \\ 
  \hline
  Power law slope dU/dM $\approx$ U$^\alpha$ & $\alpha$ & 2.4 & 2.4 \\ 
  \hline
  Dust fraction in PDRs & $\gamma$ & 0.54 &  0.54  \\ 
  \hline
  \hline
    \multicolumn{4}{c}{}\\
\hline
    \multicolumn{4}{c}{\bf No AGN emission}\\
\hline\hline
\end{tabular}}
  \caption{CIGALE modules and input parameters used for all the fits. BC03 means \protect\cite{Bruzual2003}, and the Chabrier IMF refers to \protect\cite{Chabrier2003}.}
  \label{Tab.pcigale.ini}
\end{center}
\end{table*}

\section{Results assuming a periodic star formation history}
\label{appendix:periodic}

We present the same analysis obtained in the main paper, but here, we assume a periodic SFH, that is, a series of regular bursts over the age of galaxies (Fig.~\ref{fig:MdustMstar-periodic}). The main parameters that define the SFH for this CIGALE run are listed in Tab.~\ref{Tab.pcigale.ini}. The conclusions presented in the main article could also be reached with a periodic SFH, confirming that the type of SFH does not fundamentally impact the results of the article. To also test if the wavelength range or the type of data (photometric or spectroscopic) could influence the results presented in this paper, we also present the same M$_{dust}$ vs. M$_{star}$ diagram in Fig.~\ref{fig:MdustMstar-others}: while we do not observe any meaningful differences in the top panel if we do not use the sub-mm data, the two sequences detected in Fig.~\ref{fig:MdustMstar} disappear in the bottom panel, showing that the information from the spectrum is fundamental for identifying the stardust and ISM dust sequences.

\begin{figure}
	\includegraphics[width=\columnwidth]{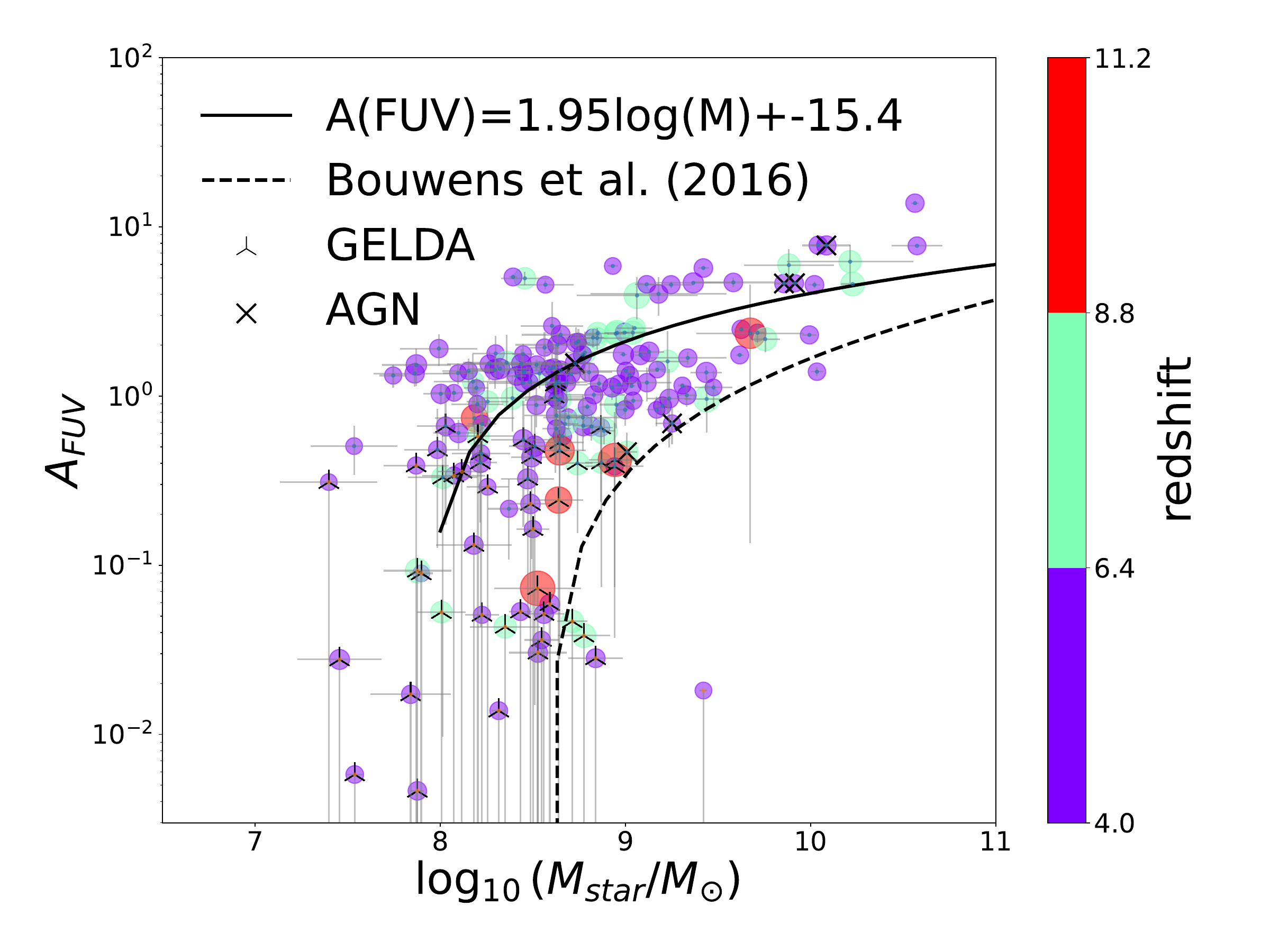}
    \caption{This figure is the same as Fig.~\ref{fig:AFUV-Mstar}. However, for this one we assume a periodic SFH. This alternate version allows to reach the same conclusion than Fig.~\ref{fig:AFUV-Mstar}, with a population of GELDAs.}
    \label{fig:AFUV-Mstar-periodic}
\end{figure}

\begin{figure}
	\includegraphics[width=\columnwidth]{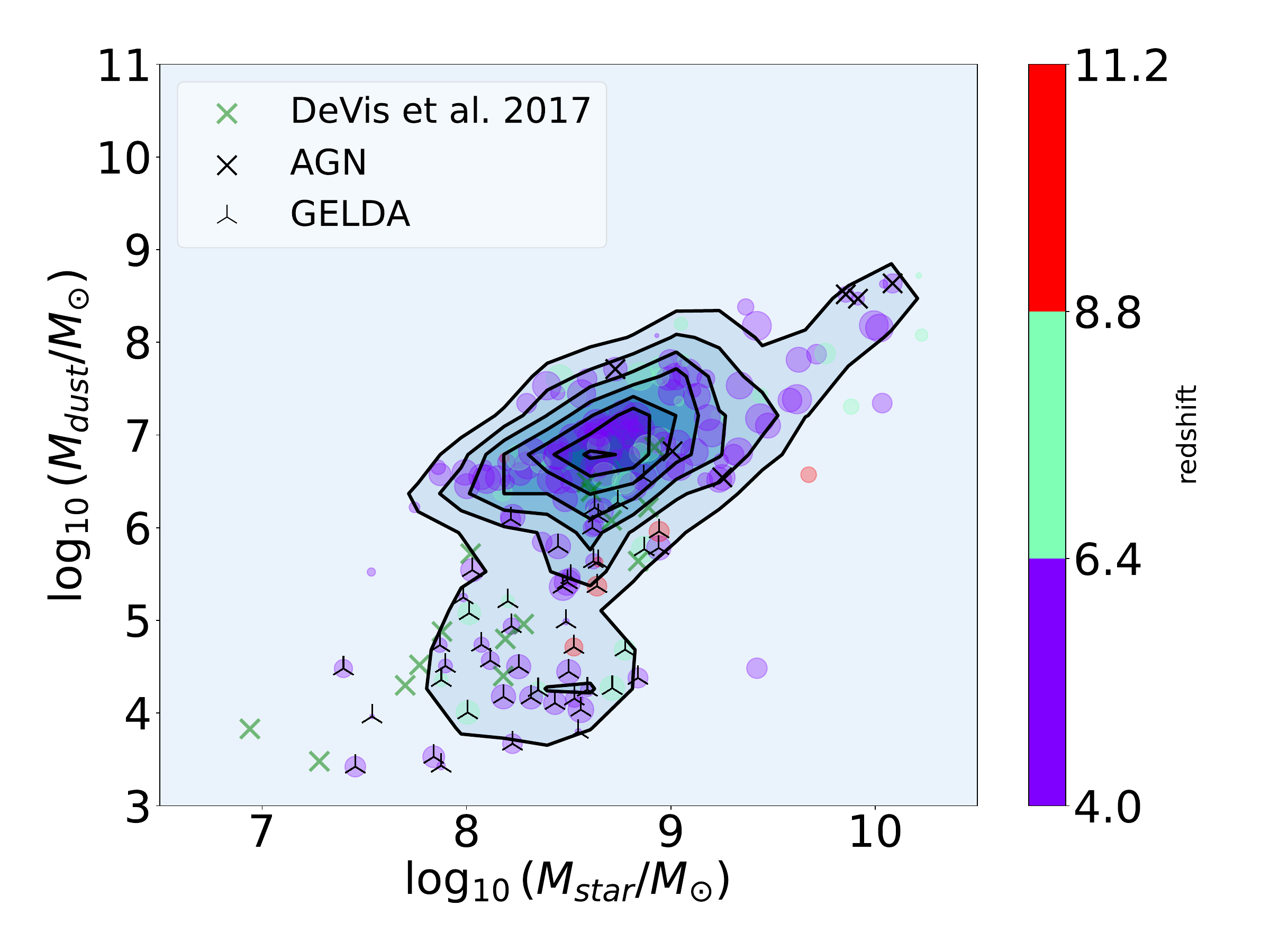}
    \caption{This figure is the same as Fig.~\ref{fig:MdustMstar}. However, for this one we assume a periodic SFH. This alternate version also allows to reach the same conclusion than Fig.~\ref{fig:AFUV-Mstar}, with a population of GELDAs.}
    \label{fig:MdustMstar-periodic}
\end{figure}

\begin{figure}
	\includegraphics[width=\columnwidth]{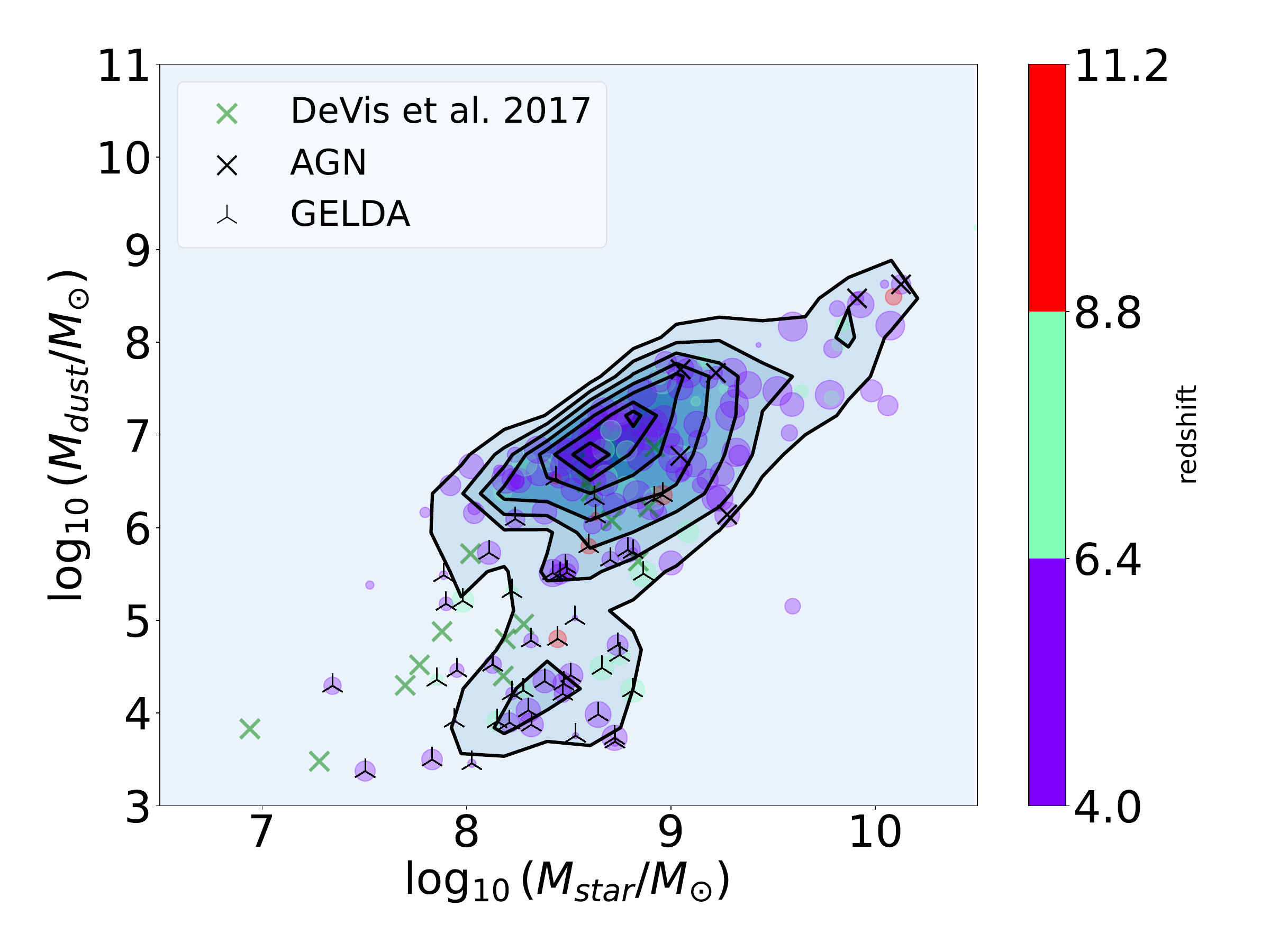}
	\includegraphics[width=\columnwidth]{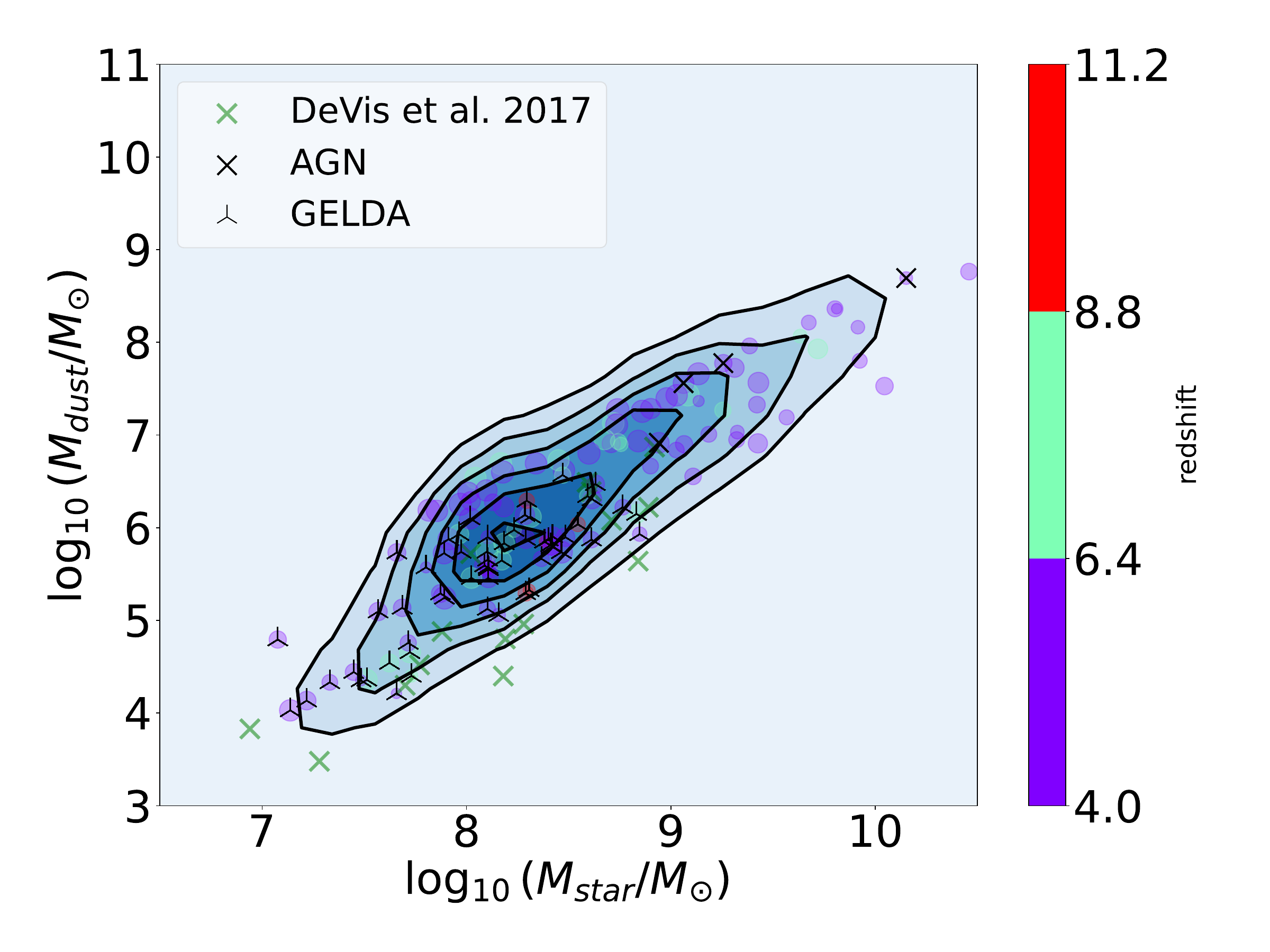}
    \caption{Test on the stability of the results with the type of data used: Top - This figure is created by fitting the spectrophotometric data, as in Fig.~\ref{fig:MdustMstar} and Fig.~\ref{fig:MdustMstar-periodic}, except that we do not use the sub-mm ones. The trend observed in this case is almost identical, which confirms the less important role of the sub-mm data in separating the two parallel sequences. Bottom - This figure is created by only fitting the photometric data, including the sub-mm ones but not the spectroscopic one. The trend observed in this case is different from when we make use of the NIRSpec spectra. We still do see a small decrease at lower stellar mass, even though the less-marked downturn suggests that without spectroscopy, some strong spectral information, and especially the line ratios, is missing. However, the photometric data still bring an information on the dust attenuation because of the UV slope $\beta_{FUV}$. The correlation of $\beta_{FUV}$ with the dust mass is much less significant, and leads to this smaller difference in dust mass, even at low stellar masses which makes the second lower sequence less prominent.}
    \label{fig:MdustMstar-others}
\end{figure}

\end{appendix}
\end{document}